\documentclass[onecolumn,amsmath,amssymb,12pt,superscriptaddress,nofootinbib]{revtex4-2}
\usepackage[T1]{fontenc}
\usepackage{lmodern}
\usepackage[english]{babel}
\usepackage{amsthm}
\usepackage{amssymb}
\usepackage{amsmath}
\usepackage{amsthm}
\usepackage[]{graphicx}
\usepackage{tensor}
\usepackage{framed}
\usepackage{framed}
\usepackage[bookmarks,linktocpage, colorlinks=true, plainpages = false, citecolor = blue,  linkcolor=blue, urlcolor = blue, filecolor = blue]{hyperref} 
\usepackage{natbib}
\usepackage{dcolumn}
\usepackage{bm}
\usepackage{float}
\usepackage{tikz}
\usetikzlibrary{decorations.pathmorphing, patterns,shapes}
\usepackage[utf8]{inputenc}
\usepackage[T1]{fontenc}
\usepackage{xcolor}
\usepackage{physics} 
\usepackage{slashed}
\usepackage{dsfont}
\usepackage{float}
\usepackage{simplewick}
\usepackage{microtype}
\usepackage{graphicx}
\usepackage{vwcol} 
\usepackage{url}
\usepackage{hyperref}
\usepackage{indentfirst}
\begin{document}
\allowdisplaybreaks
\begin{titlepage}

\title{
No-boundary solutions are robust to quantum gravity corrections
}
\author{Caroline Jonas}
\email{caroline.jonas@aei.mpg.de}
\affiliation{Max--Planck--Institute for Gravitational Physics (Albert--Einstein--Institute), 14476 Potsdam, Germany}
\author{Jean-Luc Lehners}
\email{jlehners@aei.mpg.de}
\affiliation{Max--Planck--Institute for Gravitational Physics (Albert--Einstein--Institute), 14476 Potsdam, Germany}

\begin{abstract}
\vspace{.5cm}
\noindent 
The no-boundary proposal is a theory of the initial conditions of the universe formulated in semi-classical gravity, and relying on the existence of regular (complex) solutions of the equations of motion. We show by explicit computation that regular no-boundary solutions are modified, but not destroyed, upon inclusion of expected quantum gravity corrections that involve higher powers of the Riemann tensor as well as covariant derivatives thereof. We illustrate our results  with examples drawn from string theory. Our findings provide a crucial self-consistency test of the no-boundary framework.
\end{abstract}
\maketitle
\end{titlepage}
\tableofcontents

%%%%%%%%%%%%%%%%%%%%%%%%%
%%%%%%%%%%%%%%%%%%%%%%%%%
%%%%%%%%%%%%%%%%%%%%%%%%%
%%%%%%%%%%%%%%%%%%%%%%%%%
%%%%%%%%%%%%%%%%%%%%%%%%%
%%%%%%%%%%%%%%%%%%%%%%%%%
%%%%%%%%%%%%%%%%%%%%%%%%%
%%%%%%%%%%%%%%%%%%%%%%%%%
%%%%%%%%%%%%%%%%%%%%%%%%%
%%%%%%%%%%%%%%%%%%%%%%%%%

\section{Introduction} \label{sec:intro}

The Hartle-Hawking no-boundary proposal \cite{1993AdSAC...8..223H,Hartle:1983ai} provides a theory of the quantum state of the universe. As such it is a theory of the initial conditions of the universe, meaning that it provides (relative) probabilities for different evolutions of the universe \cite{Hartle:2008ng}. The proposal is formulated in semi-classical gravity and relies on the existence of solutions of the Einstein equations that replace the big bang singularity with a smooth geometry. In Lorentzian signature it is however not possible to find a regular solution that starts out at zero size. The insight of Hartle and Hawking was that in Euclidean signature regular solutions can exist, the prototype being a 4-sphere of constant positive curvature. In the simplest case of a cosmological constant one may then think of a no-boundary geometry as a  gluing of a Euclidean onto a Lorentzian solution. Once a scalar field is added the solutions are necessarily complex, and they smoothly interpolate between Euclidean and Lorentzian signature \cite{Lyons1992}. 

There are two crucial features of no-boundary solutions, namely that they are compact and that they are regular (i.e.~Euclidean) near the big bang. Both features are necessary in order to obtain a consistent semi-classical description. However, from a quantum point of view, these two features do not commute: compactness requires specifying a vanishing initial size while regularity corresponds to specifying an initial Euclidean expansion rate. Since size and expansion rate are conjugate variables that must satisfy the uncertainty principle, both conditions cannot be imposed simultaneously. Recent work has shown that fixing a zero initial size leads to trouble \cite{Feldbrugge:2017fcc}, while one can obtain a consistent path integral definition of the no-boundary proposal when one specifies the initial expansion rate to be Euclidean \cite{DiTucci:2019dji,DiTucci:2019bui}. This construction is also supported by the analogous calculation in anti de-Sitter space, where one may use well known results in black hole thermodynamics as guidance \cite{DiTucci:2020weq}. Thus the latest understanding of the no-boundary proposal is that it should not be thought of as a sum over compact metrics, but rather as a sum over geometries of all sizes that start out as purely spatial (Euclidean) metrics. Then, as the universe grows, the signature changes to Lorentzian -- time is not present at the ``beginning'', where one only has space. The no-boundary geometry, which is both Euclidean \emph{and} compact, then arises as the dominant (saddle point) contribution to the path integral. 

The regularity of no-boundary geometries is crucial to the proposal since otherwise there is no chance that one may trust the results of semi-classical gravity. After all, gravity is non-renormalisable and one expects an eventual full theory of quantum gravity to have an effective description as general relativity augmented by a series of quantum corrections of higher order in the Riemann tensor. A singularity in the solution would imply an infinite sensitivity to such curvature corrections. But then one must wonder whether a solution with the required characteristics (regularity, finite action) still exists in the presence of the expected quantum gravity corrections. This is the topic of the present paper. 

If we were looking for solutions with constant 4-curvature, the answer would be almost trivial since terms of higher order in the Riemann tensor (even with covariant derivatives included) would have a simple structure and such corrections would be suppressed with powers of the 4-curvature (assumed to be well below the Planck scale). But realistic no-boundary solutions have varying curvature, and can be quite different from the toy model (half-sphere + de Sitter) geometry. Moreover there exist ekpyrotic no-boundary solutions which have a geometrical shape that is very different from that of inflationary instantons \cite{Battarra:2014xoa,Battarra:2014kga}. Technically, the problem may be formulated as follows: in a universe with scale factor $a(t),$ the Riemann tensor contains terms of the form 
\begin{align}
Riem \, \sim \, \frac{1}{a^2}\,, \, \frac{\dot{a}^2}{a^2}\,, \, \frac{\ddot{a}}{a}\,,
\end{align}
and thus it is not at all clear that there will be a smooth solution when $a \to 0.$ In fact, it seems that the problem will get worse when considering higher powers of the Riemann tensor\footnote{Very few works have looked into this question in the past, in particular see Hawking and Luttrell \cite{Hawking:1984ph} and Vilenkin \cite{Vilenkin1985} on quadratic gravity, and van Elst et al. on including a cubic Ricci scalar term \cite{vanElst:1994jg}.}. Nevertheless, as we will show in this paper, there exist conspiracies between the various terms in the Riemann tensor such that for a large class of theories, including all the known corrections stemming from string theory, smooth solutions continue to exist. Even when covariant derivatives are included in the correction terms, no-boundary solutions are robust to these corrections in the sense that the solutions will be modified somewhat, but their smoothness property is not endangered. This result represents an important self-consistency check of the no-boundary proposal,  as it implies that the results obtained using only the setting of semi-classical gravity will continue to hold without drastic modification in more complete theories of quantum gravity. 

The plan of this article is as follows. We will begin in section \ref{sec:noboundary} by reviewing the salient features of the no-boundary proposal that we will require. In section \ref{sec:riemannterms} we will consider all actions composed solely of Riemann terms, i.e.~terms that are scalar contractions of Riemann tensors, for metrics of closed Friedmann-Lema\^{i}tre-Robertson-Walker (FLRW) form. Then in section \ref{sec:examples} we will focus on specific extensions of general relativity and quantum gravity corrections, and see if they admit a consistent and regular no-boundary solution. Section \ref{sec:covariantderivatives} will be devoted to the study of covariant derivatives of Riemann terms, that appear in some quantum gravity corrections. Our conclusions are in section \ref{sec:conclusion}. We employ the convention that the Riemann tensor is defined as $
	R^{\lambda}_{\ \mu\alpha\nu}=\partial_{\alpha}\Gamma^{\lambda}_{\mu\nu}-\partial_{\nu}\Gamma^{\lambda}_{\mu\alpha}+\Gamma^{\beta}_{\mu\nu}\Gamma^{\lambda}_{\beta\alpha}-\Gamma^{\beta}_{\mu\alpha}\Gamma^{\lambda}_{\beta\nu}\;$
	and the Ricci tensor as $
	R_{\mu\nu}=R^{\lambda}_{\ \mu\lambda\nu}\;$.

\section{The no-boundary ansatz \label{sec:noboundary}}

The no-boundary wavefunction is a function of the (e.g.~current) spatial metric of the universe $h_{ij}$ and matter configuration $\tilde\phi,$ defined as the path integral 
\begin{align}
\Psi(h_{ij},\tilde\phi) & = \int^{h_{ij},\tilde\phi} D\phi Dg_{\mu\nu} e^{\frac{i}{\hbar}S}\,, \\
S & = \frac{1}{8\pi G}\int d^4x \sqrt{-g} \left[ \frac{R}{2} - \Lambda + \cdots \right] + \frac{1}{8\pi G}\int_{h_{ij}} d^3 y \sqrt{h}K\,,
\end{align}
where in the action the dots stand for matter contributions $\phi$ and eventual additional curvature terms. The cosmological constant is denoted by $\Lambda.$ A Gibbons-Hawking-York surface term (involving the trace of the extrinsic curvature $K$) is added on the final boundary, allowing one to fix the spatial metric there, but no such term is added at the ``no-boundary hypersurface'' so as to allow for the imposition of a momentum condition there, forcing metrics to be Euclidean near the nucleation of the universe -- for full details see \cite{DiTucci:2019bui,DiTucci:2020weq}. This path integral can then be evaluated in the saddle point approximation, with a no-boundary geometry providing the dominant contribution. In the present work we will not consider the difficult problem of defining the path integral in the presence of higher derivative terms in the action, rather we will assume that the saddle point approximation will remain valid. More to the point, we will investigate whether suitable candidates for a no-boundary saddle point geometry exist. 

It is useful to first look at the case of a closed FLRW metric in the presence of perfect fluid matter. The metric is given by
\begin{equation}
\dd s^2=-N(t)^2\dd t^2+a(t)^2\left[\dd\psi^2+\sin^2\psi\left(\dd\theta^2+\sin^2\theta\dd\phi^2\right)\right],\label{metric}
\end{equation}
where $\psi$ and $\theta$ range from $0$ to $\pi$ and $\phi$ ranges from $0$ to $2\pi$. The lapse function $N(t)$ and the scale factor $a(t)$ both only depend on time. For the fluid, we will assume a stress tensor of perfect fluid form  $T^{\mu\nu}=p(t)g^{\mu\nu}+\left(\rho(t)+p(t)\right)u^{\mu}u^{\nu}$ where $\rho(t)$ is the energy density, $p(t)$ the pressure and $u^\mu$ the 4-velocity. Then the constraint and equations of motion for general relativity plus a perfect fluid are 
\begin{align}
&\frac{\dot{a}^2}{N^2}+1=\frac{a^2}{3}\left(\Lambda+ 8\pi G\rho \right)\,, \label{eq:constraint}\\
&\frac{2\ddot{a}}{aN^2}+\frac{\dot{a}^2}{a^2N^2}+\frac{1}{a^2}-\Lambda=-8\pi G p\,, \\ & a \dot{\rho} + 3 \dot{a} (\rho + p) = 0\,. 
\end{align}
We are now looking for a solution that is regular as $a(t) \to 0$ (we will choose the origin of the time coordinate such that this coincides with $t\to 0$). From the equations above one can see that this can only be achieved if 
\begin{equation}
\dot{a}^2(t\to 0)=-N^2\quad;\quad\ddot{a}(t\to 0)=0\quad;\quad (\rho + p)(t\to 0)=0\ .\label{nobsol}
\end{equation}
This is precisely the no-boundary solution. The condition on $\dot{a}$ immediately implies that the metric is Euclidean near $t=0.$ Meanwhile, the condition on the energy density and pressure implies that near $t=0$ the only form of matter that is allowed is one which has the equation of state of a cosmological constant there. An example is a scalar field that approaches a constant value at $t=0,$ i.e.~for which $\dot\phi(t=0)=0.$ No other form of matter is allowed near the ``big bang'' (also sometimes called the South Pole of the instanton), as this would destroy the regularity of the solution. This means that for our purposes we can actually ignore matter contributions and focus only on gravitational terms.

Given that we need to focus on gravitational terms, do we need to worry mainly about anisotropies near the South Pole? To see that this is not the case, consider a Bianchi IX metric,
\begin{align}
\dd s^2_{IX}=-N^2\dd t^2+\frac{a^2}{4}&\Big[e^{\beta_++\sqrt{3}\beta_-}(\sin\psi\dd\theta-\cos\psi\sin\theta\dd\phi)^2+e^{\beta_+-\sqrt{3}\beta_-}(\cos\psi\dd\theta+\sin\psi\sin\theta\dd\phi)^2\nonumber\\
&\ +e^{-2\beta_+}(\dd\psi+\cos\theta\dd\phi)^2\Big]\,;
\end{align}
in $(t,\psi,\theta,\phi)$ coordinates, with $\theta\in[0,\pi]$, $\phi\in[0,2\pi]$ and $\psi\in[0,4\pi]$. 
Neglecting matter, the constraint and equations of motion for the Einstein-Hilbert action are
\begin{align}
& \frac{3\dot{a}^2}{a^2}-\frac{3}{4}(\dot{\beta}^2_++\dot{\beta}^2_-)-\frac{N^2}{a^2}U(\beta_{+},\beta_-) - N^2 \Lambda\,=\,0\,;\\
&\frac{\dot{a}^2}{a^2N^2}+\frac{2\ddot{a}}{aN^2}+\frac{3}{4N^2}(\dot{\beta}^2_++\dot{\beta}^2_-)-\frac{1}{3a^2}U(\beta_+,\beta_-) - \Lambda =0\,; \label{BIXeom}
\end{align}
where
\begin{equation}
U(\beta_{+},\beta_{-})=e^{-4\beta_{+}}+e^{2\beta_+-2\sqrt{3}\beta_-}+e^{2\beta_++2\sqrt{3}\beta_-}-2e^{2\beta_+}-2e^{-\beta_+-\sqrt{3}\beta_-}-2e^{-\beta_++\sqrt{3}\beta_-}\,.
\end{equation}
Close to $t=0$ the no-boundary ansatz \eqref{nobsol} again leads to a solution, provided that in addition $(\dot{\beta}^2_++\dot{\beta}^2_-)(t\to 0)=0$ and $U(\beta_+,\beta_-)(t\to 0)=-3.$ This implies that the anisotropies $\beta_+$ and $\beta_-$ are necessarily going to zero when $t\to0$. Similar arguments apply to inhomogeneities. This means that as long as a homogeneous and isotropic solution exists, there can always be other solutions which develop inhomogeneities and anisotropies away from the South Pole, while approaching the most symmetric solution at the South Pole. This will remain true when we consider more involved theories of gravity.

We conclude that close to the no-boundary point, we can focus on the isotropic and homogeneous part of the metric, i.e.~on the scale factor. To determine the existence of no-boundary solutions we will therefore make use of a Taylor series ansatz of the form
\begin{equation}
\left\lbrace
\begin{aligned}
&a(t)=a_1t+\frac{a_3}{6}t^3+\frac{a_4}{24}t^4+\frac{a_5}{120}t^5+O(t^6)\,;\\
&a_1^2=-N^2\,.
\end{aligned}
\right.\label{firstansatz}
\end{equation}
Our aim will be to see if such a series solution exists in the presence of quantum gravity corrections. Before embarking on this task, a few remarks:
\begin{enumerate}
	\item \label{complexconjugated}The regularity condition $\dot{a}^2(0)=-N^2$ leads to two complex conjugated solutions, $\dot{a}(0)= a_1 = \pm iN$. These actually correspond to the Vilenkin \cite{Vilenkin:1982de} and Hartle-Hawking \cite{Hartle:1983ai} choices. Our present work will not distinguish between the two, but for discussions of the differences see e.g.~\cite{Feldbrugge:2017kzv,Feldbrugge:2017fcc,Feldbrugge:2017mbc,Vilenkin:2018dch,Feldbrugge:2018gin}.
	\item The coefficient $a_1=\pm i N$ on its own just describes flat space. Therefore, $a(t)=a_1t$ will always be a solution of any action constructed purely from Riemann tensors. However it is not clear whether  for arbitrary actions we can have non-vanishing $a_3, a_5,\dots$ coefficients that will define a no-boundary solution regular in time.
	\item \label{remarka3}
	The coefficient $a_3$ is related to how fast the universe is expanding. This can be seen from the no-boundary solution for general relativity in the presence of a cosmological constant $\Lambda \equiv 3 H^2$, which in Euclidean time $\tau=-iNt$ is given by
	\begin{equation}
	a(\tau)=\frac{1}{H}\sin(H\tau) = \tau-\frac{1}{6}H^2\tau^3+\cdots\,.
	\end{equation}
	We recover $a_1^2=-N^2$, independently of $H$, and moreover we can see that $a_3$ is proportional to $H^2$. Therefore, for generic theories that allow solutions with different expansion rates, we should expect $a_3$ to remain a free parameter, labelling the various solutions. These solutions with different expansion rates will have different actions, and thus obtain different probabilities. In fact it is in this sense that the no-boundary proposal provides a quantum theory of initial conditions.
	\end{enumerate}

%%%%%%%%%%%%%%%%%%%%%%%%%%%%%%%%%%%%%%
%%%%%%%%%%%%%%%%%%%%%%%%%%%%%%%%%%%%%%

\section{Riemann terms \label{sec:riemannterms}}

In this section we will investigate the impact of adding terms of higher order in the Riemann tensor, without the inclusion of covariant derivatives. As explained in the previous section, we can reduce our investigation to that of the scale factor in a closed FLRW universe, with metric \eqref{metric}. In this spacetime, the only non-vanishing components of the Riemann tensor $R^{\mu\nu}_{\ \ \,\rho\sigma}$ are of the form $R^{ab}_{\ \ ab}$ and $R^{ab}_{\ \ ba}$  with $a,b=0,\dots, 3$, $a\neq b$ and no summation on $a$ and $b$ implied. Therefore all scalar contractions composed of $n$ Riemann tensors $R_{\mu\nu\rho\sigma}$ and $2n$ inverse metrics $g^{\mu\nu}$ can in this FLRW background be written as contractions of $n$ $R^{ab}_{\ \ ab}$ or $R^{ab}_{\ \ ba}$ (where $n$ can be any integer). Moreover, these 24 non-zero components have simple expressions in terms of the lapse and scale-factor functions:  $\forall\  i,j=1,2,3$ with $i\neq j$ and no summation on the indices implied,
\begin{align}
&R^{ij}_{\ \ ij}=\frac{\dot{a}^2+N^2}{a^2N^2}\equiv A_1\quad\mbox{and}\quad R^{0i}_{\ \ 0i}=\frac{\ddot{a}N-\dot{a}\dot{N}}{aN^3}\equiv A_2\,.\label{riemannterms}
\end{align}
We define a Riemann term to be any scalar combination of Riemann tensors and metric terms. As a consequence of \eqref{riemannterms}, any Riemann term can be written as a polynomial in $A_1$ and $A_2$ on a closed FLRW background. Basic examples are the Ricci scalar $R=6(A_1+A_2)$, the Ricci tensor squared $R^{\mu\nu}R_{\mu\nu}=12(A_1^2+A_1A_2+A_2^2)$ 
and the Riemann tensor squared $R^{\mu\nu\rho\sigma}R_{\mu\nu\rho\sigma}= 12(A_1^2+A_2^2)$.

%%%%%%%%%%%%%%%%%%%%%%%%%%%

\subsection{General action and constraint}

Since all Riemann terms are polynomials in $A_1$ and $A_2$, the most general action containing only such terms will take the form
\begin{align}
S&=\int\dd^4x\sqrt{-g}\cdot f\left(R_{\mu\nu\rho\sigma},g^{\alpha\beta}\right)=2\pi^2\int\dd t\ a^3 N\sum_{p_1,p_2\in\mathbb{N}^2}c_{p_1,p_2}A_1^{p_1}A_2^{p_2}\,,\label{genactionA1A2}
\end{align}
where $c_{p_1,p_2}$ is a constant depending on the precise form of $f$ for each couple $\lbrace p_1,p_2\rbrace$.

In order to later find the equations of motions, we slightly manipulate this action.
The lapse $N$ is a non-dynamical variable whose equation of motion is a constraint on the system. Therefore, given that we will work in a gauge where $N$ is constant, any term containing more than one power of $\dot{N}$ will later disappear at the level of the equations of motion. Decomposing $A_2^{p_2}$ with the Newton formula,
\begin{equation}
A_2^{p_2}=\left(\frac{\ddot{a}}{aN^2}-\frac{\dot{a}\dot{N}}{aN^3}\right)^{p_2}=\sum_{l=0}^{p_2}{p_2\choose l}\left(-\frac{\dot{a}\dot{N}}{aN^3}\right)^l\left(\frac{\ddot{a}}{aN^2}\right)^{p_2-l},
\end{equation}
the relevant part is given by the terms $l=0$ and $l=1$, so we replace
\begin{equation}
\left(\frac{\ddot{a}N-\dot{a}\dot{N}}{aN^3}\right)^{p_2}\to\left(\frac{\ddot{a}}{aN^2}\right)^{p_2-1}\left(\frac{\ddot{a}}{aN^2}-p_2\frac{\dot{a}\dot{N}}{aN^3}\right).
\end{equation}
We also rewrite
\begin{equation}
A_1^{p_1}=\left(\frac{\dot{a}^2+N^2}{a^2N^2}\right)^{p_1}=\frac{1}{a^{2p_1}}\sum_{j=0}^{p_1}{p_1\choose j}\frac{\dot{a}^{2j}}{N^{2j}}\,.
\end{equation}
The action \eqref{genactionA1A2} then reduces to
\begin{equation}
S=2\pi^2\sum_{p_1,p_2\in\mathbb{N}^2}c_{p_1,p_2}\sum_{j=0}^{p_1}{p_1\choose j}\int\dd t\left[\frac{1}{N^{2p_2-1+2j}}\frac{\dot{a}^{2j}\ddot{a}^{p_2}}{a^{2p_1+p_2-3}}-p_2\frac{\dot{N}}{N^{2p_2+2j}}\frac{\dot{a}^{2j+1}\ddot{a}^{p_2-1}}{a^{2p_1+p_2-3}}\right].\label{generalaction}
\end{equation}

We can now calculate the constraint equation  by variating the general action \eqref{generalaction} with respect to the lapse function $N(t)$. Using
\begin{equation}
\frac{\dot{N}}{N^{2p_2+2j}}=\frac{\dd}{\dd t}\left(-\frac{1}{2p_2+2j-1}\cdot\frac{1}{N^{2p_2+2j-1}}\right)\,,
\end{equation}
we can rewrite \eqref{generalaction} as
\begin{align}
S=2\pi^2\sum_{p_1,p_2}c_{p_1,p_2}\sum_{j=0}^{p_1}{p_1\choose j}\int&\frac{\dd t}{N^{2p_2+2j-1}}\bigg[\frac{\dot{a}^{2j}\ddot{a}^{p_2}}{a^{2p_1+p_2-3}}-\frac{p_2}{2p_2+2j-1}\Big\lbrace(2j+1)\frac{\dot{a}^{2j}\ddot{a}^{p_2}}{a^{2p_1+p_2-3}}\nonumber\\
&+(p_2-1)\frac{\dot{a}^{2j+1}\ddot{a}^{p_2-2}\dddot{a}}{a^{2p_1+p_2-3}}-(2p_1+p_2-3)\frac{\dot{a}^{2j+2}\ddot{a}^{p_2-1}}{a^{2p_1+p_2-2}}\Big\rbrace\bigg]\\
\equiv 2\pi^2\sum_{p_1,p_2}c_{p_1,p_2}\sum_{j=0}^{p_1}{p_1\choose j}\int&\frac{\dd t}{N^{2p_2+2j-1}}\cdot\mathcal{L}_{p_1,p_2,j}(a,\dot{a},\ddot{a},\dddot{a})\,.	
\end{align}
Varying w.r.t. the lapse then yields
\begin{equation}
\delta_N S=2\pi^2\sum_{p_1,p_2}c_{p_1,p_2}\sum_{j=0}^{p_1}{p_1\choose j}\int\dd t\left(\frac{-(2p_2+2j-1)\delta N}{N^{2p_2+2k}}\right)\cdot\mathcal{L}_{p_1,p_2,j}\ ;
\end{equation}	
so that the constraint equation of this system is
\begin{align}
0=\frac{\delta S}{\delta N}&=-2\pi^2\sum_{p_1,p_2}c_{p_1,p_2}\sum_{j=0}^{p_1}{p_1\choose j}\frac{2p_2+2j-1}{N^{2p_2+2j}}\cdot\mathcal{L}_{p_1,p_2,j}\ ;\\
\Leftrightarrow\ 0=\frac{\delta S}{\delta N}&=-2\pi^2\sum_{p_1,p_2}\frac{c_{p_1,p_2}}{N^{2p_2}}\frac{\ddot{a}^{p_2-1}}{a^{2p_1+p_2-2}}\sum_{j=0}^{p_1}{p_1\choose j}\frac{\dot{a}^{2j}}{N^{2j}}\bigg[(2j-1)(1-p_2)a\ddot{a}\label{constraint2}\\
&\quad\quad\quad\quad\quad\quad\quad\quad\quad\quad\quad\quad\quad\quad\quad-p_2(p_2-1)\frac{a\dot{a}a^{(3)}}{\ddot{a}}+p_2(2p_1+p_2-3)\dot{a}^{2}\bigg]\,.\nonumber
\end{align}
Using Newton's binomial formula,
\begin{equation}
\left\lbrace
\begin{aligned}
&\sum_{j=0}^{p_1}{p_1\choose j}\frac{\dot{a}^{2j}}{N^{2j}}=\left(\frac{\dot{a}^{2}}{N^2}+1\right)^{p_1}= a^{2p_1}A_1^{p_1}\,;\\
&\sum_{j=0}^{p_1}{p_1\choose j}j\frac{\dot{a}^{2j}}{N^{2j}}=p_1\frac{\dot{a}^2}{N^2}\left(\frac{\dot{a}^2}{N^2}+1\right)^{p_1-1}= p_1\frac{\dot{a}^2}{a^2N^2}a^{2p_1}A_1^{p_1}\,;
\end{aligned}
\right.
\end{equation}
the constraint equation \eqref{constraint2} reduces to
\begin{align}
0=\frac{\delta S}{\delta N}=2\pi^2&\sum_{p_1,p_2}c_{p_1,p_2}\bigg[2p_1(p_2-1)\frac{a\dot{a}^2}{N^2}A_2^{p_2}A_1^{p_1-1}+(1-p_2)a^3A_2^{p_2}A_1^{p_1}\nonumber\\
&\quad\quad\quad+p_2(p_2-1)\frac{a\dot{a}a^{(3)}}{N^4}A_2^{p_2-2}A_1^{p_1}-p_2(2p_1+p_2-3)\frac{a\dot{a}^2}{N^2}A_2^{p_2-1}A_1^{p_1}\bigg]\,.\label{Friedmannconstraintnotsimplified}
\end{align}
We have verified that the equation of motion for the scale factor, obtained by varying the action with respect to $a,$ is implied by the constraint equation in the sense that it can be obtained by deriving the constraint with respect to time. From now on we shall therefore work exclusively with the constraint equation \eqref{Friedmannconstraintnotsimplified}.

%%%%%%%%%%%%%%%%%%%%%%%%%%%%%%%%%%%%%%

\subsection{Order by order equations with the no-boundary ansatz}

Now we are ready to insert the no-boundary ansatz into the Friedmann constraint equation \eqref{Friedmannconstraintnotsimplified} for the general action \eqref{generalaction}. We will then analyse the resulting equations order by order in $t$. This will provide conditions the action must obey so as to admit a no-boundary solution. 

We first make the observation that the constraint equation \eqref{Friedmannconstraintnotsimplified} (hence also the equation of motion), and the no-boundary conditions \eqref{nobsol}, are all invariant under the transformation
\begin{equation}
\left\lbrace
\begin{aligned}
&t\to -t,\\
&a \to -a,
\end{aligned}
\right.
\quad\Rightarrow\quad a(-t)=-a(t)\ ;\label{aisodd}
\end{equation}
so the function $a$ must be odd in $t$. Thus all coefficients of even powers of $t$ in the Taylor expansion are zero, and the no-boundary ansatz \eqref{firstansatz} can in fact be simplified to
\begin{equation}
\left\lbrace
\begin{aligned}
&a(t)=a_1t+\frac{a_3}{6}t^3+\frac{a_5}{120}t^5+O(t^7)\,;\\
&a_1^2=-N^2\,.
\end{aligned}
\right.\label{ansatz}
\end{equation}
The fact that $a$ is an odd function of $t$ implies that for any solution $a(t)$, there will always exist a time-reversed solution, but both will have the same signature as the metric only depends on $a(t)^2$. For this second solution, the proper time runs in the opposite coordinate time direction $t$. Since there is also always a complex conjugate solution for each solution (see \ref{complexconjugated}), this makes for four solutions in total.

We start by plugging \eqref{ansatz} into the expressions for $A_1$ and $A_2,$ obtaining the expansions
\begin{equation}
\left\lbrace
\begin{aligned}
&A_1=\frac{\dot{a}^2(t)+N^2}{a^2(t)N^2}=-\frac{a_3}{a_1^3}+\frac{\big(a_3^2-a_1a_5\big)}{12a_1^4}\,t^2+\frac{\big(a_3a_5-a_1a_7\big)}{360a_1^4}\,t^4+O(t^6)\,;\\
&A_2=\frac{\ddot{a}(t)}{a(t)N^2}=-\frac{a_3}{a_1^3}+\frac{\big(a_3^2-a_1a_5\big)}{6a_1^4}\,t^2-\frac{\big(10a_3^3-13a_1a_3a_5+3a_1^2a_7\big)}{360a_1^5}\,t^4+O(t^6)\,.
\end{aligned}
\right.\label{A1A2expansion}
\end{equation}
The fact that these expansions start at order $t^{0}$ is non-trivial since $A_1$ and $A_2$ both contain powers of $a(t)$ in their denominators, so they could in principle have been singular as $t\to 0$, but this is precisely what the no-boundary solution prevents. The combination $A_2-A_1$ only starts at order $t^2$. 

Then we plug the no-boundary ansatz \eqref{ansatz} into the Friedmann constraint equation \eqref{Friedmannconstraintnotsimplified} (see appendix \ref{appendix:constrainteq}). The surprise is that even though we allow terms of arbitrary order in the Riemann tensor, all coefficients of negative powers of $t$ vanish automatically and the first non-trivial condition arises at order $t.$ In fact, at the two lowest non-trivial orders ($t$ and $t^3$) we obtain two conditions on the coefficients $c_{p_1,p_2}$:
\begin{align}
&\mbox{Order $t$ :}\quad\sum_{p_1,p_2}\frac{c_{p_1,p_2}}{N^{2P}}a_1^{4-P}a_3^{P-1}\big(p_2-p_1\big)=0\,;\label{LOcondition}\\
&\mbox{Order $t^3$ :}\quad\sum_{p_1,p_2}\frac{c_{p_1,p_2}}{N^{2P}}\,a_1^{3-P}a_3^{P-2}\,\Big(a_3^2\cdot G_3[p_1,p_2]+a_1a_5\cdot G_{5}[p_1,p_2]\Big)=0\,;\label{NNLOcondition}
\end{align}
where $P\equiv p_1+p_2$ and
\begin{align}
G_3[p_1,p_2]=\frac{1}{6}\left(p_1^2-15p_1+6-4p_2^2+12p_2\right)\;;\ 
G_5[p_1,p_2]=\frac{p_1(1-p_1)}{6}-\frac{2p_2(1-p_2)}{3}\,.
\end{align}

One way of easily satisfying the first condition \eqref{LOcondition} is by requiring that
\begin{equation}
\forall \lbrace p_1,p_2\rbrace\in\mathbb{N}^2\ ,\quad c_{p_1,p_2}=c_{p_2,p_1}\,.\label{restrictedcond}
\end{equation}
This special case in fact covers most known examples:
\begin{itemize}
	\item any term of the form $R^{n},\ \forall\,n\in\mathbb{N},$ satisfies \eqref{restrictedcond} since $R=6(A_1+A_2)$. In particular this implies that $f(R)$ theory, and hence gravity plus a scalar field, will admit a no-boundary solution.
	\item quadratic terms and all their powers since $R_{\mu\nu\rho\sigma}R^{\mu\nu\rho\sigma}=6\left(A_1^2+A_2^2\right)$ and $R_{\mu\nu}R^{\mu\nu}=12\left(A_1^2+A_1A_2+A_2^2\right)$.\\
\end{itemize}

We then turn to the second condition \eqref{NNLOcondition}.
Provided the expression factoring $a_5$ is not zero, this condition in fact determines the value of $a_5$ in terms of $a_1$ and $a_3$:
\begin{equation}
a_5\cdot\sum_{p_1,p_2}\frac{c_{p_1,p_2}}{N^{2P}}\,a_1^{3-P}a_3^{P-2}G_5[p_1,p_2]=-\frac{a_3^2}{a_1}\cdot\sum_{p_1,p_2}\frac{c_{p_1,p_2}}{N^{2P}}\,a_1^{3-P}a_3^{P-2}G_3[p_1,p_2]\,.\label{a5expr}
\end{equation}
When we are in the special case where \eqref{restrictedcond} is satisfied, we can simplify \eqref{a5expr} by symmetrising the expressions $G_3$ and $G_5$ in the exchange of $p_1$ and $p_2$, and we find
\begin{align}
&a_5=-\frac{a_3^2}{a_1}\cdot\frac{\sum_{p_1,p_2}\frac{c_{p_1,p_2}}{N^{2P}}a_1^{3-P}a_3^{P-2}\Big[4-p_1(p_1+1)-p_2(p_2+1)\Big]}{\sum_{p_1,p_2}\frac{c_{p_1,p_2}}{N^{2P}}a_1^{3-P}a_3^{P-2}\Big[p_1(p_1-1)+p_2(p_2-1)\Big]}\,.
\end{align}
At higher orders in $t$ the additionally appearing coefficients $a_7, a_9, \dots$ will be fixed in terms of the lower ones. Thus all theories of this form admit no-boundary solutions as $a \to 0,$ with $a_3$ remaining a free parameter effectively corresponding to solutions with different expansion rates. 

The single exception to this statement is the case where the left-hand side of \eqref{a5expr} vanishes, with the consequence that $a_3$ is fixed in terms of $a_1$. This corresponds to ordinary general relativity in the presence of a cosmological constant. Expanding \eqref{eq:constraint} one straightforwardly finds
\begin{equation}
a_3=-\frac{a_1^3\Lambda}{3}\ ;\ a_5=-\frac{5a_3^2}{a_1}-2a_1^2a_3\Lambda=\frac{a_1^5\Lambda^2}{9}\,; \quad  \textrm{etc}.
\end{equation}
For this theory the no-boundary solution corresponds to complexified de Sitter space with fixed expansion rate determined by the cosmological constant.

What we have done so far is to find general conditions that Riemann terms need to satisfy if they are to preserve the existence of no-boundary solutions.  In the next section we will examine specific examples of extensions of general relativity to see whether or not they fulfil these requirements. But before doing so it may be helpful, for the sake of illustration, to see what goes wrong if the condition \eqref{LOcondition} is not satisfied. Even though we do not have a covariant expression for them, let us consider actions like
\begin{equation}
\int\dd t\, a^3N A_1\;;\quad\mbox{or}\quad\int\dd t\, a^3N A_1A_2^2\;;\quad\mbox{etc,}
\end{equation}
that are in violation of \eqref{LOcondition}.
The constraint equation for the action $\int\dd t a^3N A_1$ gives
\begin{equation}
(a_1^2-N^2) t+a_1a_3 t^3+O(t^5)=0\,;
\end{equation}
so even in the presence of matter (only appearing at order $t^3$), this would imply $a_1=\pm N,$ corresponding to Minkowski spacetime rather than Euclidean space near $a=0.$ This is inconsistent with the no-boundary ansatz. Here we see that it is not enough for an approximately flat solution to exist near $a=0,$ it must be flat and Euclidean at the same time. Even this is not enough, as the next example will show: if we turn to $\int\dd t a^3N A_1A_2^2$ for instance, the constraint equation is
\begin{equation}
	\frac{2a_1a_3^2}{N^6}t+\Big(\frac{4a_1a_3a_5}{3N^6}-2a_1^3\Lambda\Big)t^3+O(t^5)=0\ ;
\end{equation}
where we have included a cosmological constant $\Lambda$ and assumed the no-boundary relation $a_1^2 = - N^2.$ At order $t$ one is forced to set $a_3$ to zero, but then at the next order the constraint cannot be satisfied. Hence this action does not admit a no-boundary solution.

Having gained a better appreciation for the non-triviality of the no-boundary regularity condition we now turn our attention to specific examples of theories containing higher orders of the Riemann tensor in the action.

%%%%%%%%%%%%%%%%%%%%%%%%%%%%%%%%%%%%%%
%%%%%%%%%%%%%%%%%%%%%%%%%%%%%%%%%%%%%%

\section{No-boundary solutions for extensions of general relativity \label{sec:examples}}

%%%%%%%%%%%%%%%%%%%%%%%%%%%%%%%%%%%%%%

\subsection{Quadratic gravity} \label{sec:qg}

The most straightforward extension of Einstein gravity is quadratic gravity, analysed in this context in \cite{Hawking:1984ph,Vilenkin1985}. It has the advantage of being a renormalisable theory of gravity  \cite{Stelle1978}, but it suffers from the presence of a ghost. Lots of efforts are being made in order to make sense of this ghost, see e.g.~\cite{Donoghue:2019fcb,Salvio:2019ewf}. Quadratic gravity has many uses, such as in Starobinsky's inflation \cite{Starobinsky:1986fx}, in asymptotic safety \cite{Codello:2006in,Einhorn:2014gfa,Salvio:2017qkx}, and it has interesting general implications near the big bang, where it automatically enforces the suppression of certain classes of anisotropies and inhomogeneities \cite{Lehners:2019ibe} (even for a big bang that is not of no-boundary type and that gives rise to a curvature singularity).

We will first consider pure quadratic gravity, where the action only contains $R^2$ terms. This theory is scale invariant and has the action
\begin{equation}
S_{\text{pure quad}}=\int\dd^4x\sqrt{-g}\left(\alpha R^2+\beta R_{\mu\nu}R^{\mu\nu}+\gamma R_{\mu\nu\rho\sigma}R^{\mu\nu\rho\sigma}\right).\label{puregrav}
\end{equation}
On closed FLRW background, we recall that
\begin{align}
R^2=36\left(A_1+A_2\right)^2\ ;\ R_{\mu\nu}R^{\mu\nu}=12\left(A_1^2+A_1A_2+A_2^2\right)\ \mbox{and}\ R_{\mu\nu\rho\sigma}R^{\mu\nu\rho\sigma}=12\left(A_1^2+A_2^2\right).\nonumber
\end{align}
In four dimensions, the Gauss-Bonnet term $\displaystyle\int\dd^4x\sqrt{-g}\,\mathcal{G}\,$  is a topological invariant and does not contribute to the dynamics. On a closed FLRW background,
\begin{equation}\mathcal{G}\equiv R^{\alpha\beta\gamma\delta}R_{\alpha\beta\gamma\delta}-4R^{\alpha\beta}R_{\alpha\beta}+R^2=24A_1A_2\,;
\end{equation} and the associated constraint equation obtained by inserting $\lbrace p_1=1,p_2=1\rbrace$ in \eqref{Friedmannconstraintnotsimplified} is automatically null.
To study the dynamics the action can therefore effectively be reduced to 
\begin{equation}
\left.S_{\text{pure quad}}\right\rvert_{reduced}=2\pi^2\int\dd t\,a^3N\,\epsilon(A_1^2+A_2^2)\quad;\quad\mbox{with}\ \epsilon=36\alpha+12\beta+12\gamma\,.\label{reducedpurequad}
\end{equation}

This time even at order $t^1$ the constraint equation is automatically satisfied because the action \eqref{reducedpurequad} is symmetric in $A_1$ and $A_2$, and therefore satisfies the condition \eqref{restrictedcond}. At next order in $t$, the constraint equation yields
\begin{equation}
\frac{\left(a_1a_3^2- a_1^2 a_5\right)}{N^4}\cdot t^3+O\left(t^5\right)=0\;,
\end{equation}
solved by $a_5=a_3^2/a_1.$ The coefficient $a_3$ is left undetermined, as expected from the scale invariance of the theory.

Next we can consider coupling quadratic gravity to ordinary general relativity,
\begin{equation}
S_{\text{quad}}=\int\dd^4x\sqrt{-g}\left(\frac{R}{16\pi G}-\frac{\Lambda}{8\pi G}+\frac{\omega}{3\sigma} R^2-\frac{1}{2\sigma}C^2+\epsilon\mathcal{G}\right),
\end{equation}
where we wrote the action in terms of the Weyl tensor $C,$ which vanishes for a FLRW metric:
\begin{align}
C_{\mu\nu\rho\sigma}C^{\mu\nu\rho\sigma}=&\,R_{\mu\nu\rho\sigma}R^{\mu\nu\rho\sigma}-2R_{\mu\nu}R^{\mu\nu}+\frac{1}{3}R^2=\,0\,;\nonumber
\end{align}
and the Gauss-Bonnet combination $\mathcal{G}$, which does not contribute to the dynamics as we just saw. Therefore the relevant part of the quadratic action to compute the dynamics on a FLRW background is 
\begin{equation}
S_{\text{quad,reduced}}=2\pi^2\int\dd t a^3N\Big[\frac{1}{8\pi G}\left(3A_1+3A_2-\Lambda\right)+\frac{12\omega}{\sigma}(A_1^2+A_2^2)\Big]\,.
\end{equation}
The constraint equation for this action is
\begin{equation}
\left(\alpha 
a_1^3 \Lambda +\frac{a_3^2 \beta
}{a_1^3}-\frac{a_5\beta }{a_1^2}+3 \alpha 
a_3\right)\cdot t^3+O\left(t^5\right)=0\,;
\end{equation}
where $\displaystyle\alpha=\frac{1}{8\pi G}$ and $\displaystyle\beta=\frac{12\omega}{\sigma}$.
The no-boundary solution is 
\begin{equation}
a_5=\frac{a_3^2}{a_1}+\frac{\alpha}{\beta}\big(a_1^5\Lambda+3a_1^2a_3\big)\,;\label{solquadgrav1}
\end{equation}
valid when $\alpha\sim\beta$ or $\alpha\ll\beta$. Then $a_3$ is left undetermined.

When $\alpha\gg\beta$, the solution is instead
\begin{equation}
a_3\,=\, \frac{-3a_1^3\pm\sqrt{9a_1^6-4\frac{\beta}{\alpha}a_1^6\Lambda+4\frac{\beta^2}{\alpha^2}a_1a_5}}{2\beta/\alpha}\ \xrightarrow{\alpha\gg\beta} \left\lbrace
\begin{aligned}
&-\frac{a_1^3\Lambda}{3}+O\big(\beta/\alpha\big)\ ;\\
&-3a_1^3\frac{\alpha}{\beta}+\frac{a_1^3\Lambda}{3}+O\big(\beta/\alpha\big)\ .
\end{aligned}
\right.\label{solquadgrav2}
\end{equation}
The first branch corresponds to the Einstein-Hilbert solution, while the second branch is not physical as it gives a solution with curvature $a_3$ bigger than the Planck scale ($\alpha$), and a non smooth limit $\beta\to0$. The second branch arises due to the presence of higher derivatives in the action, and is associated with the new scalar degree of freedom (for a detailed discussion of the properties of the scalar, see e.g.~\cite{Alvarez-Gaume:2015rwa}).

%%%%%%%%%%%%%%%%%%%%%%%%%%%%%%%%%%%%%%%%%%%%%%%%%%%%%%%%%%
\subsection{Heterotic string theory}

The low-energy effective theory from heterotic string theory is the \textit{Einstein -- Maxwell -- axion -- dilaton} gravity containing a dilaton field $\phi$, gauge fields $F$ (Maxwell) and a 3-form $H$ (axion), see e.g.~\cite{METSAEV1987385,Ohta:2012ih}. At first order in the inverse string tension $\alpha^{\prime}$, an S-matrix calculation in heterotic string theory leads to the effective Einstein frame action \cite{Ohta:2012ih}
\begin{equation}
S_{\text{heterotic}}=\frac{1}{2\kappa^2_{D}}\int\dd^Dx\sqrt{-g}\bigg(R-\frac{1}{2}(\partial\phi)^2+\frac{\alpha^{\prime}}{8}e^{-\phi/2}\Big(\mathcal{G}+\frac{3}{16}(\partial\phi)^4\Big)-V(\phi) +\cdots\bigg)\,;\label{heteroticaction}
\end{equation}
where we have assumed that the compactification has led to a potential $V(\phi)$ for the dilaton (in general we may expect additional terms). Note that, as discussed in section \ref{sec:noboundary}, the axion $H$ and the gauge fields $F$ have been  consistently set to zero. If additional scalar fields arise due to the compactification, then these will behave analogously to the dilaton, so that we may use the dilaton as a stand-in for all of the scalars. In the gravitational sector, the first correction in $\alpha^{\prime}$ is given by the Gauss-Bonnet combination. Because of the dilaton dependent prefactor, it is not a topological invariant this time, and we must include its effects. The constraint reads
\begin{align}
\frac{\delta}{\delta N}\Big(\mathcal{L}_{\text{heterotic}}\Big)=&\,\frac{\delta}{\delta N}\Big(6a^3N(A_1+A_2)\Big)-a^3\Big[\frac{\dot{\phi}^2}{2}-\frac{\alpha^{\prime}}{128}e^{-\phi/2}\dot{\phi}^4+V(\phi)\Big]\nonumber\\
&+3\alpha^{\prime}e^{-\phi/2}\frac{\delta}{\delta N}\Big(a^3NA_1A_2\Big)-\frac{3\alpha^{\prime}}{2}\dot{\phi}\,e^{-\phi/2}\frac{\dot{a}a^2}{N^2}A_1=0\,,\label{eomphi1}
\end{align}
where the second line follows from
\begin{equation}
\frac{\delta}{\delta N}\Big(a^3N\,A(N,\dot{N},t)\,B(t)\Big)=B\,\frac{\delta \big(a^3NA\big)}{\delta N}-\dot{B}\,\frac{\partial \big(a^3NA\big)}{\partial\dot{N}}\ \mbox{for}\  A\equiv \mathcal{G}\ \mbox{and}\ B\equiv e^{-\phi/2}\,.
\end{equation}
Equation \eqref{eomphi1} is odd under the transformation $t\to -t$, $a\to -a$ and $\phi\to\phi$. We will also need the equation of motion for the scalar $\phi,$ which is given by
\begin{align}
\nabla^2\phi-\frac{\alpha^{\prime}}{16}e^{-\phi/2}\Big(\mathcal{G}+3(\nabla_{\mu}\phi)(\nabla_\nu\phi)(\nabla^{\mu}\nabla^{\nu}\phi)+\frac{3}{2}\nabla^2\phi(\partial\phi)^2-\frac{9}{16}\big(\partial\phi\big)^4\Big)- V_{,\phi}=0\,.
\end{align}
On a closed FLRW background and for a homogeneous field $\phi(t)$ this translates into
\begin{equation}
\ddot{\phi}-\frac{3\alpha^{\prime}}{16}e^{-\phi/2}\Big(8A_1A_2+\frac{3}{2}\dot{\phi}^2\ddot{\phi}-\frac{3}{16}\dot{\phi}^4\Big)-V_{,\phi}=0\,.\label{eomphi2}
\end{equation}
This equation \eqref{eomphi2} is even under the transformation $t\to -t$, $a\to-a$ and $\phi\to\phi$.

Now we look for Taylor series solutions to equations \eqref{eomphi1} and \eqref{eomphi2} around $t=0$. From the transformation rules of the equations of motion \eqref{eomphi1} and \eqref{eomphi2} under $t\to -t$, $a\to-a$ and $\phi\to\phi$, we know that $a$ must be an odd function of time, while $\phi(t)$ must be even:
\begin{equation}
\left\lbrace
\begin{aligned}
&a=a_1 t+\frac{a_3}{6} t^3+\frac{a_5}{120}t^5+\dots\\
&\phi(t)=\phi_0+\frac{\phi_2}{2} t^2+\frac{\phi_4}{24} t^4+\dots
\end{aligned}
\right.\label{phiansatz}
\end{equation}
This is already enough to realise that $\phi$ will be constant at first order in time close to the no-boundary point $t\to 0$. When plugging \eqref{phiansatz} in the constraint equation \eqref{eomphi1} and expanding in orders of $t$, the leading order gives
\begin{equation}
-\frac{3a_1e^{-\phi_0/2}\alpha^{\prime}}{2N^4}(a_1^2+N^2)\phi_2t+O(t^3)=0\,;
\end{equation}
that is solved by the usual no-boundary solution $a_1^2=-N^2$. Then we turn to the equation of motion for $\phi$ \eqref{eomphi2} where  at leading order we find
\begin{equation}
\phi_2-\frac{3a_3^2}{2a_1^6}\alpha^{\prime}e^{-\phi_0/2}-V_{,\phi}(\phi_0)+O(t^2)=0\,.
\end{equation}
This equation fixes $\phi_2$ as a function of $\phi_0$, $a_1$ and $a_3$. Implementing this solution for $\phi_2$, the next order of the constraint equation gives us a cubic equation for $a_3$ in terms of $a_1$ and $\phi_0$:
\begin{align}
\bigg[-\frac{9a_3^3}{4a_1^8}\alpha^{\prime\, 2}e^{-\phi_0}+6a_3\Big(1-e^{-\phi_0/2}\frac{\alpha^{\prime}V_{,\phi}(\phi_0)}{4a_1^2}\Big)-a_1^3 V(\phi_0)\bigg]t^3+O(t^5)=0\,.
\end{align}

We conclude that the heterotic string action \eqref{heteroticaction} possesses a family of no-boundary solutions, this time usefully labelled by $\phi_0,$ the dilaton value at the South Pole.

%%%%%%%%%%%%%%%%%%%%%%%%%%%%%%%%%%%%%%
\subsection{General relativity as an Effective Field Theory} \label{sec:eft}

We just saw that the leading correction stemming from the heterotic string is a combination of quadratic terms in the Riemann tensor. More generally, when considering an effective field theory treatment of general relativity, in addition to the pure gravitational terms we would also expect the presence of new couplings between the gravitational terms and matter terms \cite{Donoghue:1994dn}. Of greatest interest in the present context is the coupling between gravity and scalar fields. We will not be able to perform an exhaustive treatment of such couplings, but the first non-trivial couplings serve as an indication that no obstruction to the existence of no-boundary solutions will come from such terms. To see this, consider the effective theory of gravity and a scalar field up to to fourth order in derivatives, 
\begin{align}
S_\textrm{eff}=\int\dd^4x\sqrt{-g}\,\Big[&\frac{1}{16\pi G}(R-2\Lambda)+\frac{1}{2}g^{\mu\nu}\partial_\mu\phi\partial_\nu\phi-V(\phi)+c_1R^2+c_2R_{\mu\nu}R^{\mu\nu}\nonumber\\
&+(d_1R^{\mu\nu}+d_2Rg^{\mu\nu})\partial_\mu\phi\partial_\nu\phi+d_3RV(\phi)+\cdots\Big]\,,
\end{align}
for arbitrary coefficients $c_1, c_2, d_1, d_2, d_3$. On our closed FLRW background and for a homogeneous scalar field, up to total derivatives this action reduces to
\begin{align}
S_{\textrm{eff}}=2\pi^2\int\dd t\,a^3N\bigg[&\frac{1}{8\pi G}\big(3A_1+3A_2-\Lambda\big)-\frac{\dot{\phi}^2}{2N^2}-V(\phi)+12(3c_1+c_2)(A_1^2+A_2^2)\nonumber\\
&-\big(3d_1A_2+6d_2(A_1+A_2)\big)\frac{\dot{\phi}^2}{2N^2}+6d_3(A_1+A_2)\,V(\phi)\bigg]\,.
\end{align}
By variation we can calculate the equations of motion, the scalar field equation being 
\begin{align}
\frac{a^3}{N}\bigg[&\ddot{\phi}+3H\dot{\phi}-N^2V_{,\phi}+6d_1\Big(A_2(\ddot{\phi}+2H\dot{\phi})+\frac{a^{(3)}}{aN^2}\dot{\phi}\Big)\nonumber\\
&+12d_2\Big(\ddot{\phi}(A_1+A_2)+H\dot{\phi}(A_1+4A_2)+\frac{a^{(3)}}{aN^2}\dot{\phi}\Big)+6d_3(A_1+A_2)N^2V_{,\phi}\bigg]=0\,;\label{eomphi}
\end{align}
while that for $N$ (the constraint equation) is
\begin{align}
a^3\bigg[&\frac{3A_1-\Lambda}{8\pi G}+\frac{\dot{\phi}^2}{2N^2}-V(\phi)-12(3c_1+c_2)\Big(\frac{2\dot{a}^2}{a^2N^2}A_1-\frac{2\dot{a}a^{(3)}}{a^2N^4}+\big(\frac{\ddot{a}}{aN^2}-\frac{\dot{a}^2}{a^2N^2}\big)^2\Big)\nonumber\\
&+6d_1\Big(\frac{\dot{\phi}^2}{N^2}\big(\frac{\ddot{a}}{aN^2}-\frac{\dot{a}^2}{a^2N^2}\big)-\frac{\dot{a}}{aN}\frac{\dot{\phi}\ddot{\phi}}{N^3}\Big)+6d_2\Big(\frac{\dot{\phi}^2}{N^2}A_1+\frac{2\ddot{a}}{aN^2}\frac{\dot{a}^2}{N^2}-\frac{2\dot{a}}{aN}\frac{\dot{\phi}\ddot{\phi}}{N^3}\Big)\nonumber\\
&+6d_3\Big(A_1V(\phi)+\frac{\dot{a}}{aN}\frac{\dot{\phi}}{N}V_{,\phi}\Big)\bigg]=0\,.
\end{align}
These equations transform only by an overall sign under $t\to -t$, $a\to-a$ and $\phi\to\phi.$ Thus it is again appropriate to use the ansatz Eq.~\eqref{phiansatz}, for which the equations of motion reduce to
\begin{align}
\textrm{EoM for $\phi$ :}\quad&\frac{6 a_1 t}{N^3} \left(a_1^2+N^2\right) \left(4 d_2 \phi_2+d_3 N^2 V_{,\phi}(\phi_0)\right)+O\left(t^3\right)=0\,;\\
\textrm{EoM for $N$:}\quad&-\frac{12}{a_1N^4t}\cdot(3 c_1+c_2)\left(3 a_1^4+2 a_1^2 N^2-N^4\right) +\frac{\left(a_1^2+N^2\right)t}{a_1^2N^4}\cdot\bigg[\frac{3 a_1^3 N^2}{8\pi G}\nonumber\\
&-18a_1^2a_3(3c_1+c_2)-2a_3(3c_1+c_2)N^2+6a_1^3d_3N^2V(\phi_0)\bigg]+O\left(t^3\right)=0\,;
\end{align}
which are consistent with $a_1^2=-N^2$. Higher orders in $t$ fix higher coefficients $a_3$, $a_5$, $\phi_2,\dots$ in terms of $\phi_0$ and $a_1$.
For example the next order of the $\phi$ equation gives
\begin{equation}
\frac{t^3}{N}\cdot\Big(4\phi_2(a_1^3-6a_3(d_1+4d_2))+a_1^2(a_1^3+12a_3d_3)V_{,\phi}(\phi_0)\Big)+O(t^5)=0\,;
\end{equation}
that one can use to fix the value of $\phi_2$. The crucial point is that even in the presence of higher derivative couplings, the scalar field does not diverge near the South Pole, but approaches a constant, just as for minimal coupling. Hence, even though we cannot explicitly check all possible higher derivative couplings, we may assume with some confidence that such couplings do not yield any divergences. We will thus focus our attention on pure gravitational terms.  

We should also mention that an effective treatment of general relativity leads to the appearance of non-local terms, e.g.~terms of the form $\displaystyle\int \sqrt{-g} R\frac{1}{\Box}R$ \cite{Donoghue:1994dn}. These terms may have interesting implications in cosmology, see e.g.~\cite{Calzetta:1986pj,Donoghue:2014yha,Belgacem:2017cqo}. When expanding such terms around a specific background, one obtains an infinite series with terms containing more and more derivatives. Below we will investigate some specific correction terms containing derivatives (see section \ref{sec:covariantderivatives}), but because of technical limitations we cannot make any definite statement about large or infinite numbers of derivatives. We must therefore leave this interesting question for future work. 

%%%%%%%%%%%%%%%%%%%%%%%%%%%%%%%%%%%%%%

\subsection{Type II string theory in D=10 spacetime dimensions}

The low-energy effective action, obtained by looking at quantum corrected amplitudes for four-graviton scattering\footnote{To obtain a more general action, one also has to consider five- and six-graviton scatterings in the action, see e.g \cite{Richards:2008jg} where it is shown that at one-loop level, five-graviton scattering only matters at order $\alpha^{\prime\,6}$.  Our aim is however not to be exhaustive so we will keep to the four-graviton scattering here.} in type II string theory in $D=10$ dimensions order by order in $\alpha^{\prime}$, reads \cite{Green:2010wi,Fleig:2015vky}
\begin{equation}
S=\int\dd^{D}x\sqrt{-G}\left(R+(\alpha^{\prime})^3\mathcal{E}^{(D)}_{(0,0)}\mathcal{R}^4+(\alpha^{\prime})^5\mathcal{E}^{(D)}_{(1,0)}\nabla^4\mathcal{R}^4+(\alpha^{\prime})^6\mathcal{E}^{(D)}_{(0,1)}\nabla^6\mathcal{R}^4+\dots\right);\label{staction}
\end{equation}
	where $G$ is the determinant of the metric in $D$ dimensions, while $\mathcal{E}^{(D)}_{(p,q)}$ are coefficient functions that depend on the compactification. General compactifications imply the presence of additional curvature terms (along the lines discussed above) and scalars (discussed in section \ref{subsec:typeII}) as well as numerous gauge fields which we can set to zero (cf. the discussion in section \ref{sec:noboundary}). Here we will focus on the $\alpha^{\prime 3}$ type II correction to Einstein gravity \eqref{staction} which is given by the $\mathcal{R}^4$ term, a special combination of four Riemann tensors defined as\footnote{Again this will be modified when considering five-graviton scattering by the addition of a $\epsilon_8\epsilon_8R^4$ term. As we will discuss on next page, this kind of term will not be relevant for our analysis and we can safely ignore it.}
\begin{equation}
\mathcal{R}^4=t_8^{ijklmnpq}t_8^{abcdefgh}R_{ijab}R_{klcd}R_{mnef}R_{pqgh}\,.
\end{equation}
$t_8$ is a special 8-rank tensor whose explicit expression can be found in \cite{Green:2012pqa} (chapter 9, Appendix A) to be:
\begin{align}
t^{ijklmnpq}=-\frac{1}{2}&\epsilon^{ijklmnpq}\nonumber\\
-\frac{1}{2}&\Big[\left(\delta^{ik}\delta^{jl}-\delta^{il}\delta^{jk}\right)\left(\delta^{mp}\delta^{nq}-\delta^{mq}\delta^{np}\right)+\left(\delta^{km}\delta^{ln}-\delta^{kn}\delta^{lm}\right)\left(\delta^{pi}\delta^{qj}-\delta^{pj}\delta^{qi}\right)\nonumber\\
&+\left(\delta^{im}\delta^{jn}-\delta^{in}\delta^{jm}\right)\left(\delta^{kp}\delta^{lq}-\delta^{kq}\delta^{lp}\right)\Big]\nonumber\\
+\frac{1}{2}&\Big[\delta^{jk}\delta^{lm}\delta^{np}\delta^{qi}+\delta^{jm}\delta^{nk}\delta^{lp}\delta^{qi}+\delta^{jm}\delta^{np}\delta^{qk}\delta^{li}+\mbox{ 45 more terms}\nonumber\\
&\mbox{obtained by antisymmetrizing on the pairs } ij,\ kl,\ mn\ \mbox{and}\  pq\ \Big].
\end{align}
The quantity $\mathcal{R}^4$ is therefore a Riemann term, so we can determine if it will admit a no-boundary solution by simply looking at its structure in terms of $A_1$ and $A_2$ and see if it meets condition \eqref{LOcondition}. We start by computing the explicit structure of $\mathcal{R}^4$ in terms of Riemann tensors with the xAct package \cite{xACT}:
\begin{align}
\mathcal{R}^4=12(R_{abcd}R^{abcd})^2&+6R^{abcd}R_{ab}^{\ \ ij}(4R_{ij}^{\ \ kl}R_{cdkl}-R_{ic}^{\ \ kl}R_{jdkl})-12R_{abij}R_{cdkl}R^{abci}R^{djkl}\nonumber\\
&+\frac{3}{2}R_{abij}R^{acid}R^{jl}_{\ \ ck}R^{bk}_{\ \ dl}+\frac{3}{4}R_{abij}R^{acid}R_{ckdl}R^{bkjl}\nonumber\\
+\epsilon^{ijklmnpq}R_{ij}^{\ \ ab}&\left[2R_{klef}R_{mn}^{\ \ \ ef}R_{pqab}-\frac{1}{2}R_{kl}^{\ \ ef}R_{mnae}R_{pqbf}-\frac{1}{2}R_{klae}R^{\ \ \ fe}_{mn}R_{pqbf}\nonumber\right.\\
&\left.+2R_{kl}^{\ \ ef}R_{mnab}R_{pqef}-\frac{1}{2}R_{klae}R_{mnbf}R_{pq}^{\ \ ef}+2R_{klab}R_{mn}^{\ \ \ ef}R_{pqef}\right]\nonumber\\
+\frac{1}{4}\epsilon^{ijklmnpq}&\epsilon^{efghabcd}R_{ijab}R_{klcd}R_{efmn}R_{ghpq}.\label{totalR4}
\end{align}

We must be aware that these expressions are originally valid only in 10 dimensions (and an analogous structure is also expected in 11-dimensional supergravity, since the low-energy type II theories are related to 11-dimensional supergravity via circle compactifications, see e.g.~\cite{Green:1997as}). When going down to 4 dimensions, there will be new fields (and different associated terms) appearing through the compactification, when indices point in the internal dimensions. These gauge fields and scalars will depend on the details of the compactification. However, as discussed in section \ref{sec:noboundary}, we expect gauge field to be zero and scalar fields constant at the no-boundary point. Therefore the only part of \eqref{totalR4} that we are really interested in is the one where all indices point in the (four) external spacetime dimensions. But then all the terms containing an 8 rank tensor $\epsilon$ are set to zero, and we are left with
\begin{align}
\left.\mathcal{R}^{4}\right\rvert_{4d, \mbox{\scriptsize{truncated}}}=&\ 12(R_{\mu\nu}^{\ \ \ \rho\sigma}R^{\mu\nu}_{\ \ \rho\sigma})^2+6R^{\mu\nu}_{\ \ \rho\sigma}R_{\mu\nu}^{\ \ \ \xi\eta}(4R_{\xi\eta}^{\ \ \kappa\lambda}R^{\rho\sigma}_{\ \ \kappa\lambda}-R_{\xi\ \ \lambda}^{\ \rho \kappa}R_{\eta\ \kappa}^{\ \sigma\ \lambda})\nonumber\\
&-12R_{\mu\nu}^{\ \ \ \xi\eta}R^{\rho\sigma}_{\ \ \kappa\lambda}R^{\mu\nu}_{\ \ \rho \xi}R_{\sigma\eta}^{\ \ \kappa\lambda}+\frac{3}{2}R^{\mu\nu}_{\ \ \xi\eta}R_{\mu\rho}^{\  \ \ \xi\sigma}R^{\eta\ \rho}_{\ \lambda\ \kappa}R_{\nu\ \sigma}^{\ \kappa\ \lambda}\nonumber\\
&+\frac{3}{4}R_{\mu\nu}^{\ \ \ \xi\eta}R^{\mu\rho}_{\ \  \xi\sigma}R_{\rho\kappa}^{\ \ \sigma\lambda}R^{\nu\kappa}_{\ \ \ \eta\lambda}\label{truncated}\,;
\end{align}
where $\mu,\nu,\rho,\sigma,\xi,\eta,\kappa,\lambda$ are now spacetime indices running from $\lbrace 0,\dots,3\rbrace$. This expression \eqref{truncated} is now ready to be expressed in terms of $A_1$ and $A_2$. Using \eqref{riemannterms}, we compute that on this background all the terms of expression \eqref{truncated} can be written in terms of two quantities that we denote $\mathcal{R}_1$ and $\mathcal{R}_2$:
\begin{align}
12(R_{\mu\nu}^{\ \ \ \rho\sigma}R^{\mu\nu}_{\ \ \rho\sigma})^2&=12^3\left(A_1^4+2A_1^2A_2^2+A_2^4\right)\label{i1}\equiv 12\mathcal{R}_1\,;\\
24R^{\mu\nu}_{\ \ \rho\sigma}R_{\mu\nu}^{\ \ \ \xi\eta}R_{\xi\eta}^{\ \ \kappa\lambda}R^{\rho\sigma}_{\ \ \kappa\lambda}&=8\vdot 12^2\left(A_1^4+A_2^4\right)\equiv 24 \mathcal{R}_2\,;\\
-6R^{\mu\nu}_{\ \ \rho\sigma}R_{\mu\nu}^{\ \ \ \xi\eta}R_{\xi\ \ \lambda}^{\ \rho \kappa}R_{\eta\ \kappa}^{\ \sigma\ \lambda}&=-12^2(A_1^4+A_2^4)=-3\mathcal{R}_2\,;\\
-12R_{\mu\nu}^{\ \ \ \xi\eta}R^{\rho\sigma}_{\ \ \kappa\lambda}R^{\mu\nu}_{\ \ \rho \xi}R_{\sigma\eta}^{\ \ \kappa\lambda}&=-12^2\vdot 4\left(A_1^4+A_1^2A_1^2+A_2^4\right)=-2\mathcal{R}_1-6\mathcal{R}_2\,;\\
\frac{3}{2}R^{\mu\nu}_{\ \ \xi\eta}R_{\mu\rho}^{\  \ \ \xi\sigma}R^{\eta\ \rho}_{\ \lambda\ \kappa}R_{\nu\ \sigma}^{\ \kappa\ \lambda}&=9\left(3A_1^4+2A_1^2A_2^2+3A_2^4\right)=\frac{1}{16}\mathcal{R}_1+\frac{3}{8}\mathcal{R}_2\,;\\
\mbox{and finally}\quad&\nonumber\\
\frac{3}{4}R_{\mu\nu}^{\ \ \ \xi\eta}R^{\mu\rho}_{\ \  \xi\sigma}R_{\rho\kappa}^{\ \ \sigma\lambda}R^{\nu\kappa}_{\ \ \ \eta\lambda}&=18\left(A_1^4+2A_1^2A_2^2+A_2^4\right)=\frac{1}{8}\mathcal{R}_1
\,.
\end{align}	
Therefore the expression \eqref{truncated} reads
\begin{align}
\left.\mathcal{R}^{4}\right\rvert_{4d, \mbox{\scriptsize{truncated}}}
=&\frac{163}{16}\mathcal{R}_1+\frac{123}{8}\mathcal{R}_2=1467 (A_1^2+A_2^2)^2+738 (A_1^4+A_2^4)\,.\label{finalr4truncated}
\end{align}
The quantities $\mathcal{R}_1$ and $\mathcal{R}_2$ are both symmetric under the exchange of $A_1$ and $A_2$, so they satisfy the condition \eqref{restrictedcond}. Therefore, the $\mathcal{R}^4$ term satisfies the leading order condition \eqref{LOcondition}, and will admit a no-boundary solution.

It might look a bit astonishing that this very complicated scalar combination of four Riemann tensors has such a simple expression in terms of $A_1$ and $A_2$, that is moreover symmetric in the exchange of $A_1$ and $A_2$. This might lead us to think that this could be a general property of any scalar combination of Riemann tensors, but if we look at the two following combinations:
\begin{align}
R_{\mu\nu}^{\ \ \ \rho\sigma}R^{\mu\xi}_{\ \ \rho\sigma}R_{\xi\kappa}^{\ \ \nu\lambda}R^{\alpha\kappa}_{\ \ \alpha\lambda}=&\ 48A_1^4+36A_2^4+48A_1A_2^3+24A_1^3A_2+60A_1^2A_2^2\,;\label{counterexample1}\\
\mbox{and}\quad\nonumber&\\
R^{\mu\nu}_{\ \ \xi\eta}R_{\mu\rho}^{\,\ \ \xi\sigma}R^{\rho\kappa}_{\ \ \nu\lambda}R_{\sigma\kappa}^{\ \ \,\eta\lambda}=&\ 12(A_1^4+A_1^2A_2^2+A_2^4)+12A_1(A_1^3+A_1A_2^2+2A_2^3)\,;\label{counterexample2}
\end{align}
we see that they are both not symmetric under the exchange of $A_1$ and $A_2$. However, they still satisfy the leading order condition \eqref{LOcondition}, and therefore admit a no-boundary solution.

We may conclude that known Riemann terms stemming from string theory have a structure that allows for no-boundary solutions. What is more, all of the covariant Riemann terms that we have investigated allow for no-boundary solutions. It would of course be very interesting if one could prove a general result in this direction. The next orders in $\alpha^{\prime}$ of the type II string theory \eqref{staction} are not Riemann terms anymore, but rather involve covariant derivatives acting on Riemann tensors. Unfortunately, it is not possible to treat covariant derivative terms as systematically as we treated Riemann terms, because they depend on higher and higher time derivatives of the scale factor $a$. We will therefore study them on a case by case basis, starting with the easiest expressions and ending with the first string theory covariant derivative term, written schematically as $\nabla^4\mathcal{R}^4$ in \eqref{staction}. 

%As we will see, the difficulty will go crescendo, and we won't study the next order $\nabla^6\mathcal{R}^4$ in this work, as we don't specifically search for a precise solution (which is anyway out of reach at present time), but rather aim at proving that a no-boundary solution can in general exist in many different types of extensions to General Relativity.

%%%%%%%%%%%%%%%%%%%%%%%%%%%%%%%%%%%%%%
%%%%%%%%%%%%%%%%%%%%%%%%%%%%%%%%%%%%%%
\section{Covariant derivatives of Riemann terms\label{sec:covariantderivatives}}

When covariant derivatives enter the game, it is even less trivial that their contributions to the constraint equation will still admit consistent and regular solutions. Indeed we have seen that Riemann terms are linear combinations of $A_1$ and $A_2$, and these quantities only start at order $t^{0}$. Therefore, when acting on them with time derivatives, there is no risk of ending up with negative powers of $t$, that could bring singularities. But the covariant derivative is also composed of the Christoffel symbol part: $\nabla\vdot\sim\partial\vdot+\ \Gamma\vdot$.  The non zero Christoffel symbols are schematically
\begin{equation}
g^{ki}  \Gamma^{0}_{ij}\sim\frac{\dot{a}}{aN^2}\quad;\quad\Gamma^{i}_{j0}\sim\frac{\dot{a}}{a}\quad\mbox{and}\quad\Gamma^{i}_{jk}\sim 1\ ;\label{christoffels}
\end{equation}
(by $\sim$ we indicate only the time dependence, not the angular dependence).
The quantity $\dot{a}/a\sim t^{-1}$ is singular, and we can fear that covariant derivatives introduce singularities into the constraint equations. Therefore we need to check term by term the existence of regular solutions in the covariant derivative terms that we need.

First consider again the transformation
\begin{equation}
\left\lbrace
\begin{aligned}
&t\to -t\,,\\
&a \to -a\,.
\end{aligned}
\right.\label{odd}
\end{equation}
On a closed FLRW background, if we consider the action 
\begin{equation}
S=\int\dd t\,a^3N\ \mathcal{L}\ ;
\end{equation}
then the constraint equation of this action will be
\begin{equation}
\frac{\delta}{\delta N}\Big(a^3N\mathcal{L}\Big)\equiv\frac{\partial\big(a^3N\mathcal{L}\big)}{\partial N}-\frac{\dd}{\dd t}\bigg[\frac{\partial\big(a^3N\mathcal{L}\big)}{\partial \dot{N}}\bigg]+\dots\ =0\ .
\end{equation}
This constraint equation will be odd under the transformation \eqref{odd} only if $\mathcal{L}$ is even under this same transformation. Now $A_1$ and $A_2$ are even under this transformation, hence such are all Riemann terms. Because the FLRW metric doesn't contain any mixed term $g_{0i}$, time derivatives will always come in pairs. The Christoffel symbols \eqref{christoffels} with one $0$ index are odd under \eqref{odd} and will also always come in pairs or with one time derivative. Therefore all covariant derivatives of Riemann terms will be even under this transformation, and their constraint equation odd. Thus we may keep using the reduced no-boundary ansatz \eqref{ansatz} instead of the full ansatz \eqref{firstansatz}.

By studying terms with up to four covariant derivatives acting on Riemann terms, we will encounter expressions with up to four derivatives acting on $a$. To ease the upcoming expressions, we therefore define
\begin{align}
&A_3\equiv\frac{a^{(3)}}{aN^3}-\frac{\dot{a}\ddot{N}}{aN^4}-\Big(\frac{3\dot{N}}{N^2}+\frac{\dot{a}}{aN}\Big)A_2\ ;\\
&A_4\equiv\frac{a^{(4)}}{aN^4}-\frac{\dot{a}N^{(3)}}{aN^5}-\frac{6\dot{N}}{N^2}A_3-\Big(\frac{6\dot{a}\dot{N}}{aN^3}+\frac{3\dot{N}^2}{N^4}+\frac{4\ddot{N}}{N^3}\Big)A_2-A_2^2\ .
\end{align}
The calculations involving covariant derivatives are rather lengthy, so we are not going to display them entirely here. Rather, we will explicitly show the simplest example that arises when two covariant derivatives act on one Riemann tensor, and relegate the results of lengthier calculations to the appendix. Our focus will be on terms of the form $\nabla^4R^4$.
%%%%%%%%%%%%%%%%%%%%%%%%%%%%%%%%%%%%%%

\subsection{An explicit example: two covariant derivatives acting on one Riemann tensor}

The following quantity is a scalar term where two covariant derivatives act on one Riemann tensor:
\begin{equation}
\mathcal{A}\equiv\laplacian{R}=-6\bigg(A_4+\frac{3\dot{a}}{aN}A_3+2A_2(A_2-A_1)\bigg)\,.\label{lapR}
\end{equation}
We can directly observe that $\mathcal{A}$ is a total derivative, so its constraint equation will be null. We will however derive this result explicitly for illustrative purposes.

To compute the constraint equation of $\mathcal{A}$ we need to compute those of the terms $A_4$ and $\displaystyle\frac{\dot{a}}{aN}A_3$, or more precisely, of the actions
\begin{equation}
S_{A_4}=\int\dd t\,a^3N A_4\quad\mbox{and}\quad S_{\dot{a}A_3}=\int\dd t\,a^3N\frac{\dot{a}}{aN}A_3\,.
\end{equation}
In a closed FLRW background, the constraint equation for the action $S_{A_4}$ is
\begin{align}
0&=\frac{\partial(a^3NA_4)}{\partial N}-\frac{\dd}{\dd t}\bigg[\frac{\partial(a^3NA_4)}{\partial\dot{N}}\bigg]+\frac{\dd^2}{\dd t^2}\bigg[\frac{\partial(a^3NA_4)}{\partial\ddot{N}}\bigg]-\frac{\dd^3}{\dd t^3}\bigg[\frac{\partial(a^3N\,A_4)}{\partial N^{(3)}}\bigg]\nonumber\\
&\equiv\frac{\delta}{\delta N}\Big[a^3NA_4\Big]\,.\label{deltaN}
\end{align}
We make the whole derivation explicitly for this first case:\footnote{In this paper, it is always implicitly understood that the following expressions are evaluated at constant lapse $N$, so that we can drop all terms containing more than one power of a derivative of $N$.}
\begin{equation}
a^3N\,A_4=\frac{a^2a^{(4)}}{N^3}-\frac{a\ddot{a}^2}{N^3}-\frac{6a^2a^{(3)}\dot{N}}{N^4}+\frac{2a\dot{a}\ddot{a}\dot{N}}{N^4}-\frac{4a^2\ddot{a}\ddot{N}}{N^4}-\frac{a^2\dot{a}N^{(3)}}{N^4};
\end{equation}
\begin{equation}
\Rightarrow\quad\left\lbrace
\begin{aligned}
\frac{\partial(a^3N\,A_4)}{\partial N}&=-\frac{3}{N^4}\Big(a^2a^{(4)}-a\ddot{a}^2\Big)\,;\\
\frac{\dd}{\dd t}\bigg[\frac{\partial(a^3N\,A_4)}{\partial \dot{N}}\bigg]
&=\frac{1}{N^4}\Big(-6a^2a^{(4)}+2\dot{a}^2\ddot{a}+2a\ddot{a}^2-10a\dot{a}a^{(3)}\Big)\,;\\
\frac{\dd^2}{\dd t^2}\bigg[\frac{\partial(a^3N\,A_4)}{\partial \ddot{N}}\bigg]
&=\frac{1}{N^4}\Big(-4a^2a^{(4)}-8\dot{a}^2\ddot{a}-8a\ddot{a}^2-16a\dot{a}a^{(3)}\Big)\,;\\
\frac{\dd^3}{\dd t^3}\bigg[\frac{\partial(a^3N\,A_4)}{\partial N^{(3)}}\bigg]
&=\frac{1}{N^4}\Big(-a^2a^{(4)}-12\dot{a}^2\ddot{a}-6a\ddot{a}^2-8a\dot{a}a^{(3)}\Big)\,.
\end{aligned}
\right.
\end{equation}
So using \eqref{deltaN} we find that the constraint equation for the action $S_{A_4}$ is
\begin{align}
\frac{\delta}{\delta N}\Big[a^3NA_4\Big]&=\frac{1}{N^4}\Big(2\dot{a}^2\ddot{a}-a\ddot{a}^2+2a\dot{a}a^{(3)}\Big)\,.\label{block1}
\end{align}
We use exactly the same procedure for all coming terms, but only display the final results. For the action $S_{\dot{a}A_3}$, we find the constraint equation to be
\begin{equation}
\frac{\delta}{\delta N}\bigg[a^3N\frac{\dot{a}}{aN}A_3\bigg]=\frac{1}{N^4}\Big(-2a\dot{a}a^{(3)}+a\ddot{a}^2-2\dot{a}^2\ddot{a}\Big)\,.\label{block2}
\end{equation}
The only missing piece to get the constraint equation for $\laplacian{R}$ \eqref{lapR} is the $A_2(A_2-A_1)$ term. This one is a simple $A_1^{p_1}A_2^{p_2}$ term, so we read off its contribution from \eqref{Friedmannconstraintnotsimplified}:
\begin{align}
\frac{\delta}{\delta N}\bigg[a^3NA_2\big(A_2-A_1\big)\bigg]
&=\frac{1}{N^4}\Big(2a\dot{a}a^{(3)}-a\ddot{a}^2+2\dot{a}^2\ddot{a}\Big)\,.
\end{align}
The constraint equation for $\mathcal{A}$ is therefore
\begin{align}
\delta\mathcal{A}\equiv\frac{\delta}{\delta N}\Big[a^3N\mathcal{A}\Big]&=-6\bigg[\frac{\delta}{\delta N}\Big[a^3NA_4\Big]+3\frac{\delta}{\delta N}\Big[a^3N\frac{\dot{a}}{aN}A_3\Big]+2\frac{\delta}{\delta N}\Big[a^3NA_2\big(A_2-A_1\big)\Big]\bigg]\nonumber\\
&=\ 0\,.
\end{align}
which is the expected result since this term is a total derivative.

\subsection{General recipe}

Using the straightforward method presented in the previous subsection, we can compute all possible covariant derivatives terms. However, we can ease our life even more by decomposing the calculations further.
Assume we know the constraint equations for the two actions
\begin{equation}
S_{A}=\int\dd t\,a^3N\,A\quad\mbox{and}\quad S_{B}=\int\dd t\, a^3N\, B\ ,
\end{equation}
where $A$ and $B$ are functions of $a$, $N$ and their time derivatives. Then the constraint equation for the action
\begin{equation}
S_{A\cdot B}=\int\dd t\, a^3N\, A\cdot B\,,
\end{equation}
will be given by
\begin{align}
\frac{\delta}{\delta N}\bigg[a^3N\,A\cdot B\bigg]=&\ A\cdot\frac{\delta}{\delta N}\Big[a^3N\,B\Big]+B\cdot\frac{\delta}{\delta N}\Big[a^3N\,A\Big]-a^3A\cdot B\nonumber\\
&\,-\dot{A}\cdot\bigg[\frac{\partial(a^3N\,B)}{\partial\dot{N}}-2\frac{\dd}{\dd t}\Big(\frac{\partial(a^3N\,B)}{\partial\ddot{N}}\Big)+3\frac{\dd^2}{\dd t^2}\Big(\frac{\partial(a^3NB)}{\partial N^{(3)}}\Big)\bigg]\nonumber\\
&\,+\ddot{A}\bigg[\frac{\partial(a^3N\, B)}{\partial\ddot{N}}-3\frac{\dd}{\dd t}\Big(\frac{\partial(a^3NB)}{\partial N^{(3)}}\Big)\bigg]-A^{(3)}\frac{\partial (a^3NB)}{\partial N^{(3)}}\nonumber\\
&\,-\dot{B}\cdot\bigg[\frac{\partial(a^3N\,A)}{\partial\dot{N}}-2\frac{\dd}{\dd t}\Big(\frac{\partial(a^3N\,A)}{\partial\ddot{N}}\Big)+3\frac{\dd^2}{\dd t^2}\Big(\frac{\partial(a^3NA)}{\partial N^{(3)}}\Big)\bigg]\nonumber\\
&\,+\ddot{B}\bigg[\frac{\partial(a^3N\, A)}{\partial\ddot{N}}-3\frac{\dd}{\dd t}\Big(\frac{\partial(a^3NA)}{\partial N^{(3)}}\Big)\bigg]-B^{(3)}\frac{\partial (a^3NA)}{\partial N^{(3)}}\,.\label{ABconstraint}
\end{align}
This assumes that the highest derivative of $N$ on which $A$ and $B$ depend is of third order, as it will be the case in this work. It is however trivial to extend \eqref{ABconstraint} to include higher orders.

Using equation \eqref{ABconstraint} enables us to build iteratively the constraint equations of more and more involved expressions of $A_1$, $A_2$, $A_3$ and $A_4$. To illustrate this, suppose we want to compute the constraint equations of the four following covariant expressions:
\begin{equation}
\begin{aligned}
&\mathcal{B}_1\equiv(\nabla_{\mu}R_{\alpha\beta\gamma\delta})(\nabla^{\mu}R^{\alpha\beta\gamma\delta})\,;\quad\mathcal{B}_2\equiv(\nabla_{\mu}R_{\alpha\beta})(\nabla^{\mu}R^{\alpha\beta})\,;\\
&\mathcal{B}_3\equiv(\nabla_{\mu}R)(\nabla^{\mu}R)\quad\quad\mbox{and}\ \quad \mathcal{B}_4\equiv(\nabla_{\mu}R^{\mu}_{\ \, \alpha\beta\gamma})(\nabla_{\nu}R^{\nu\alpha\beta\gamma})\,;\label{Bterms}
\end{aligned}
\end{equation}
that are expressed in terms of the quantities $A_1$, $A_2$ and $A_3$ as
\begin{align}
\mathcal{B}_1=\,&-12\Big[A_3^2+\frac{8\dot{a}^2}{a^2N^2}(A_2-A_1)^2\Big]\label{nabla2R23}\,;\\
\mathcal{B}_2=\,&-12\Big[A_3^2+\frac{2\dot{a}}{aN}(A_2-A_1)A_3+\frac{6\dot{a}^2}{a^2N^2}(A_2-A_1)^2\Big]\,;\label{nabla2R22}\\
\mathcal{B}_3=\,&-36\Big[A_3^2+\frac{4\dot{a}}{aN}(A_2-A_1)A_3+\frac{4\dot{a}^2}{a^2N^2}(A_2-A_1)^2\Big]\,;\label{nabla2R21}\\
\mathcal{B}_4=\,&-6\Big[2(A_2-A_1)\frac{\dot{a}}{aN}+A_3\Big]^2\,.
\end{align}
Then we just need to compute the constraint equations for the two actions 
\begin{equation}
S_{A_3}=\int\dd t\, a^3NA_3\quad\mbox{and}\quad S_{\dot{a}(A_2-A_1)}=\int\dd t\, a^3N\frac{\dot{a}}{aN}\big(A_2-A_1\big)\,;\label{basicblocks}
\end{equation}
and then combine them using \eqref{ABconstraint}.\footnote{Notice also that other terms like $R^{\mu\nu}(\laplacian{R_{\mu\nu}})$ or $R(\laplacian{R})$ can be obtained from these $\mathcal{B}$ terms \eqref{Bterms} by integrating by parts, since two terms differing by a total derivative lead to the same constraint equation.}

The general recipe we apply to compute the constraint equations of all covariant derivative terms is therefore
\begin{enumerate}
	\item Decompose the expression in terms of $A_1$, $A_2$, $A_3$ and $A_4$.\footnote{This is only valid for terms where at most four covariant derivatives are acting on Riemann tensors.}
	\item Find the basic blocks needed to build each terms in this expression (e.g.~\eqref{basicblocks} in the previous example), and compute their constraint equation.
	\item Use the formula \eqref{ABconstraint} (iteratively if needed) to combine the basic blocks and get the complete constraint equation for the initial covariant expression.
	\item Plug in the no-boundary ansatz \eqref{ansatz}. This step is commutative with the previous one.
\end{enumerate}

Using this method, we computed the constraint equations of all the $\mathcal{B}$ terms \eqref{Bterms} as well as those of the following terms where four covariant derivatives act on two Riemann tensors (see Appendix \ref{appendix:BandC}):
\begin{equation}
\begin{aligned}
&\mathcal{C}_1\equiv\laplacian{R_{\alpha\beta\gamma\delta}}\,\laplacian{R^{\alpha\beta\gamma\delta}}\ ;\ &\mathcal{C}_2\equiv\laplacian{R_{\alpha\beta}}\,\laplacian{R^{\alpha\beta}}\ ;\quad\quad\quad\  &\mathcal{C}_3\equiv\nabla^2R\,\nabla^2R\ ;\\
&\mathcal{C}_4\equiv\nabla_{\mu}\nabla_{\nu}R_{\alpha\beta\gamma\delta}\nabla^{\mu}\nabla^{\nu}R^{\alpha\beta\gamma\delta}\ ;\,&\mathcal{C}_5\equiv\nabla_{\mu}\nabla_{\nu}R_{\alpha\beta}\nabla^{\mu}\nabla^{\nu}R^{\alpha\beta}\ ;\ \ \, &\mathcal{C}_6\equiv\nabla_{\mu}\nabla_{\nu}R\,\nabla^{\mu}\nabla^{\nu}R\ .
\end{aligned}\label{Cterms}
\end{equation}
Remarkably, all the constraint equations of these expressions only start at order $t^3$, although we could expect them to start at order $t^{-1}$, and are therefore not singular. This peculiar feature will continue to hold for the cases of four derivatives acting on four Riemann tensor that we are now going to address.

\subsection{Four covariant derivatives acting on four Riemann tensors\label{subsec:d4R4}}

We are now ready to evaluate the contributions to the constraint equation stemming from the $\nabla^4\mathcal{R}^4$ terms (these terms are discussed in more detail in \cite{Green:2010kv}, see also \cite{Bossard:2014aea}). We once again consider the truncated part of $\mathcal{R}^4$, expressed in terms of the two quantities $\mathcal{R}_1$ and $\mathcal{R}_2,$
\begin{align}
\left.\mathcal{R}^{4}\right\rvert_{4d, \mbox{\scriptsize{truncated}}}
=&\frac{163}{16}\mathcal{R}_1+\frac{123}{8}\mathcal{R}_2\,;
\end{align}
with
\begin{align}
&\mathcal{R}_1=(R_{\alpha\beta\gamma\delta}R^{\alpha\beta\gamma\delta})^2\quad\mbox{and}\quad\mathcal{R}_2=R^{\alpha\beta}_{\ \ \,\gamma\delta}R_{\alpha\beta}^{\ \ \,\epsilon\zeta}R_{\epsilon\zeta}^{\ \ \eta\theta}R^{\gamma\delta}_{\ \ \eta\theta}\,.
\end{align}
There are three types of terms that one can write and that are inequivalent using integration by parts when four covariant derivatives act on four Riemann tensors:\footnote{The $R$ here does not refer to the Ricci scalar but is a schematic way of writing the Riemann tensor without bothering about the indices.}
\begin{equation}
(\nabla R)^4\quad;\quad(\nabla^2R)^2R^2\quad\mbox{and}\quad(\nabla^2R)(\nabla R)^2 R\,.
\end{equation} 
For these three types, we will construct all possible independent terms where the four Riemann tensors are either $\mathcal{R}_1$ or $\mathcal{R}_2$.
\paragraph{Type 1: $(\nabla R)^4$ terms.}
These terms can all be written as linear combinations of the four following terms:
\begin{equation}
\begin{aligned}
\mathcal{D}_1&\equiv\Big(\nabla_\mu R_{\alpha\beta\gamma\delta}\;\nabla^\mu R^{\alpha\beta\gamma\delta}\Big)^2\ ;\quad &\mathcal{D}_2\equiv\Big(\nabla_{\mu}R_{\alpha\beta\gamma\delta}\nabla_{\nu}R^{\alpha\beta\gamma\delta}\nabla^{\mu}R_{\epsilon\zeta\eta\theta}\nabla^{\nu}R^{\epsilon\zeta\eta\theta}\Big)\ ;\\
\mathcal{D}_3&\equiv\Big(\nabla_{\mu}R^{\alpha\beta}_{\,\ \ \gamma\delta}\nabla^{\mu}R_{\alpha\beta}^{\ \ \ \epsilon\zeta}\nabla_{\nu}R_{\epsilon\zeta}^{\ \ \eta\theta}\nabla^{\nu}R_{\eta\theta}^{\ \ \gamma\delta}\Big)\ ;\ &\mathcal{D}_4\equiv\Big(\nabla_{\mu}R^{\alpha\beta}_{\,\ \ \gamma\delta}\nabla_{\nu}R_{\alpha\beta}^{\ \ \ \epsilon\zeta}\nabla^{\mu}R_{\epsilon\zeta}^{\ \ \eta\theta}\nabla^{\nu}R_{\eta\theta}^{\ \ \gamma\delta}\Big)\ ;
\end{aligned}\label{Dterms}
\end{equation}
that can be expressed in terms of $A_1$, $A_2$ and $A_3$ (see Appendix \ref{appendix:EandFconstraints}).
Computing their contributions to the constraint equation requires the computation of the following constraint equations:
\begin{align}
&\Delta_1\equiv\frac{\delta}{\delta N}\bigg[a^3N\, A_3^4\bigg]=\frac{32(a_3^2-a_1a_5)^3}{9
	a_1^{15}}t^3+O\left(t^5\right)\ ;\\ 
&\Delta_2\equiv\frac{\delta}{\delta N}\bigg[a^3N\frac{\dot{a}^4}{a^4N^4}(A_2-A_1)^4\bigg]=\frac{(a_1a_5-a_3^2)^3}{72
	a_1^{15}}\cdot t^3+O\left(t^5\right)\ ;\\
&\Delta_3\equiv\frac{\delta}{\delta N}\bigg[a^3N\,\frac{\dot{a}^2}{a^2N^2}A_3^2(A_2-A_1)^2\bigg]=O\left(t^5\right)\ ;\\ &\Delta_4\equiv\frac{\delta}{\delta N}\bigg[a^3N\frac{\dot{a}^3}{a^3N^3}A_3\big(A_2-A_1\big)^3\bigg]=\frac{(a_1a_5-a_3^2)^3}{36
	a_1^{15}}\cdot t^3+O\left(t^5\right)\,.
\end{align}
Combining these, we get the contributions to the constraint coming from the four $\mathcal{D}$ terms that are displayed in Appendix \ref{appendix:EandFconstraints}. Let us stress here that up to order $t^3$, these four terms have the same structure involving the combination $a_3^2-a_1a_5$,
\begin{equation}
\delta\mathcal{D}_i=\alpha_i\frac{(a_3^2-a_1a_5)^3}{a_1^{15}}t^3+O(t^5)\,;
\end{equation}
where $\alpha_i$ are numerical factors.
\paragraph{Type 2: $(\nabla^2R)^2R^2$ terms}
In this case we can construct 8 different independent expressions:
\begin{equation}
\begin{aligned}
&\mathcal{E}_1\equiv\laplacian{R_{\alpha\beta\gamma\delta}}\big(\laplacian{R^{\alpha\beta\gamma\delta}}\big)R_{\epsilon\zeta\eta\theta}R^{\epsilon\zeta\eta\theta}\,;&\ \mathcal{E}_2\equiv\nabla_{\mu}\nabla_{\nu}(R_{\alpha\beta\gamma\delta})\nabla^{\mu}\nabla^{\nu}(R^{\alpha\beta\gamma\delta})R_{\epsilon\zeta\eta\theta}R^{\epsilon\zeta\eta\theta}\ ;\\
&\mathcal{E}_3\equiv\big((\laplacian{R_{\alpha\beta\gamma\delta}})R^{\alpha\beta\gamma\delta}\big)^2\quad\quad\quad\quad\ \ ;&\ \mathcal{E}_4\equiv\nabla_{\mu}\nabla_{\nu}(R_{\alpha\beta\gamma\delta})\nabla^{\mu}\nabla^{\nu}(R_{\epsilon\zeta\eta\theta})R^{\alpha\beta\gamma\delta}R^{\epsilon\zeta\eta\theta}\ ;\\
&\mathcal{E}_5\equiv\laplacian{R^{\alpha\beta}_{\ \ \,\gamma\delta}}\big(\laplacian{R_{\alpha\beta}^{\ \ \,\epsilon\zeta}}\big)R_{\epsilon\zeta}^{\ \ \eta\theta}R^{\gamma\delta}_{\ \ \eta\theta}\ \,;& \ \mathcal{E}_6\equiv\nabla_{\mu}\nabla_{\nu}(R^{\alpha\beta}_{\ \ \,\gamma\delta})\nabla^{\mu}\nabla^{\nu}(R_{\alpha\beta}^{\ \ \,\epsilon\zeta})R_{\epsilon\zeta}^{\ \ \eta\theta}R^{\gamma\delta}_{\ \ \eta\theta}\ \ ;\\
&\mathcal{E}_7\equiv\laplacian{R^{\alpha\beta}_{\ \ \,\gamma\delta}}\big(\laplacian{R_{\epsilon\zeta}^{\ \ \eta\theta}}\big)R_{\alpha\beta}^{\ \ \,\epsilon\zeta}R^{\gamma\delta}_{\ \ \eta\theta}\;\;;&\ \mathcal{E}_8\equiv\nabla_{\mu}\nabla_{\nu}(R^{\alpha\beta}_{\ \ \,\gamma\delta})\nabla^{\mu}\nabla^{\nu}(R_{\epsilon\zeta}^{\ \ \eta\theta})R_{\alpha\beta}^{\ \ \,\epsilon\zeta}R^{\gamma\delta}_{\ \ \eta\theta}\ \ .
\end{aligned}
\end{equation}
These are expressed in terms of the quantities $A_1$, $A_2$, $A_3$, $A_4$ and are displayed in Appendix \ref{appendix:EandFconstraints}.
\paragraph{Type 3: $(\nabla^2 R)(\nabla R)^2 R$ terms}
The possible terms constructed from $\mathcal{R}_1$ and $\mathcal{R}_2$ are:
\begin{equation}
\begin{aligned}
\mathcal{F}_1\equiv&(\laplacian{R_{\alpha\beta\gamma\delta}})R^{\alpha\beta\gamma\delta}\nabla_{\mu}R_{\epsilon\zeta\eta\theta}\nabla^{\mu}R^{\epsilon\zeta\eta\theta}\ ; &\mathcal{F}_2\equiv(\nabla_{\mu}\nabla_{\nu}R_{\alpha\beta\gamma\delta})R^{\alpha\beta\gamma\delta}\nabla^{\mu}R_{\epsilon\zeta\eta\theta}\nabla^{\nu}R^{\epsilon\zeta\eta\theta}\ ;\\	
\mathcal{F}_3\equiv&(\laplacian{R_{\alpha\beta\gamma\delta}})R_{\epsilon\zeta\eta\theta}\nabla_{\mu}R^{\alpha\beta\gamma\delta}\nabla^{\mu}R^{\epsilon\zeta\eta\theta}\ ; &\mathcal{F}_4\equiv(\nabla_{\mu}\nabla_{\nu}R_{\alpha\beta\gamma\delta})R_{\epsilon\zeta\eta\theta}\nabla^{\mu}R^{\alpha\beta\gamma\delta}\nabla^{\nu}R^{\epsilon\zeta\eta\theta}\ ;\\
\mathcal{F}_5\equiv&\big(\laplacian{R^{\alpha\beta}_{\ \ \,\gamma\delta}}\big)R_{\alpha\beta}^{\ \ \,\epsilon\zeta}\nabla_{\mu}R_{\epsilon\zeta}^{\ \ \eta\theta}\nabla^{\mu}R^{\gamma\delta}_{\ \ \eta\theta}\ ; &\mathcal{F}_6\equiv\nabla_{\mu}\nabla_{\nu}(R^{\alpha\beta}_{\ \ \,\gamma\delta})R_{\alpha\beta}^{\ \ \,\epsilon\zeta}\nabla^{\mu}R_{\epsilon\zeta}^{\ \ \eta\theta}\nabla^{\nu}R^{\gamma\delta}_{\ \ \eta\theta}\ \ ;\\
\mathcal{F}_7\equiv&\big(\laplacian{R^{\alpha\beta}_{\ \ \,\gamma\delta}}\big)R_{\epsilon\zeta}^{\ \ \eta\theta}\nabla_{\mu}R_{\alpha\beta}^{\ \ \,\epsilon\zeta}\nabla^{\mu}R^{\gamma\delta}_{\ \ \eta\theta}\ ; &\mathcal{F}_8\equiv\nabla_{\mu}\nabla_{\nu}(R^{\alpha\beta}_{\ \ \,\gamma\delta})R_{\epsilon\zeta}^{\ \ \eta\theta}\nabla^{\mu}R_{\alpha\beta}^{\ \ \,\epsilon\zeta}\nabla^{\nu}R^{\gamma\delta}_{\ \ \eta\theta}\ \ .	
\end{aligned}
\end{equation}
Again they can be expressed in terms of $A_1$, $A_2$, $A_3$ and $A_4$, see Appendix \ref{appendix:EandFconstraints}.\\

%\paragraph{Contributions to the constraint from $\mathcal{E}$ and $\mathcal{F}$ terms}
To compute the contribution to the constraint equation stemming from $\mathcal{E}$ and $\mathcal{F}$ terms, we will need to compute those of the following basic expressions:
\begin{align*}
&\gamma_{1}=A_1^2 A_2^2 (A_2-A_1)^2\ ;\quad\quad\ \ \,\gamma_{2}=A_2^4 (A_2-A_1)^2\ ;\quad\quad\quad\quad\ \,\gamma_{3}=A_1 A_2^3
(A_2-A_1)^2\ ;\quad\quad\\
&\gamma_{4}=A_1^2 A_4^2\  ;\quad\ \ \quad\quad\quad\quad\quad\quad\gamma_{5}=A_2^2 A_4^2\ ;\quad\quad\quad\quad\quad\quad\quad\quad\gamma_{6}=A_1
A_2^2 (A_2-A_1) A_4\  ;\quad\quad\\
&\gamma_{7}=A_2 A_3^2 A_4\ ;\ \quad\quad\quad\quad\quad \,\quad\gamma_{8}=A_1 A_2 (A_2-A_1)
A_3^2\ ;\quad\quad\ \ \gamma_{9}=\frac{ \dot{a}}{a N}A_1^2 A_2 (A_2-A_1) A_3\  ;\quad\quad\\
&\gamma_{10}=\frac{\dot{a}}{a N}A_2^3 (A_2-A_1) A_3 \  ;\quad\ \gamma_{11}=\frac{\dot{a}}{a N}A_1 A_2^2 (A_2-A_1)
A_3 \ ;\ \ \gamma_{12}=\frac{\dot{a}}{a N}A_1^2 A_3 A_4 \  ;\quad\quad\\
&\gamma_{13}=\frac{\dot{a}}{a N}A_1 A_2 A_3 A_4\ ;\quad\quad\quad\ \gamma_{14}=\frac{ \dot{a}}{a
	N}A_2^2 A_3 A_4\ ;\quad\quad\quad\quad\quad\gamma_{15}=\frac{\dot{a}}{a N}A_1 A_3^3 \ ;\quad\quad\\
&\gamma_{16}=\frac{
	\dot{a}}{a N}A_2 A_3^3\ ;\quad\quad\quad\quad\ \ \quad\gamma_{17}=\frac{\dot{a}^2}{a^2 N^2}A_1^2 A_2 (A_2-A_1)^2 \ ;\ \ \gamma_{18}=\frac{\dot{a}^2}{a^2 N^2}A_1 A_2^2 (A_2-A_1)^2 \  ;\quad\quad\\
&\gamma_{19}=\frac{\dot{a}^2}{a^2 N^2}A_2^3 (A_2-A_1)^2
\  ;\ \quad\gamma_{20}=\frac{\dot{a}^2}{a^2 N^2}A_1^2 A_3^2 \  ;\quad\quad\quad\quad\quad\ \gamma_{21}=\frac{ \dot{a}^2}{a^2 N^2}A_1 A_2 A_3^2\quad ;\quad\quad\\
&\gamma_{22}=\frac{\dot{a}^2}{a^2
	N^2}A_2^2 A_3^2 \  ;\quad\quad\quad\quad\quad \gamma_{23}=\frac{\dot{a}^2}{a^2 N^2}A_1^2 (A_2-A_1) A_4 \  ;\ \ \ \gamma_{24}=\frac{\dot{a}^2}{a^2 N^2}A_1 A_2 (A_2-A_1) A_4 \  ;\quad\quad\\
&\gamma_{25}=\frac{\dot{a}^2}{a^2 N^2}A_2^2 (A_2-A_1)
A_4 \ ;\ \ \gamma_{26}=\frac{\dot{a}^3}{a^3 N^3}A_1^2 (A_2-A_1) A_3 \ ;\ \ \ \gamma_{27}=\frac{\dot{a}^3}{a^3 N^3}A_1 A_2 (A_2-A_1) A_3 \ ;\quad\quad\\
&\gamma_{28}=\frac{\dot{a}^3}{a^3 N^3}A_2^2
(A_2-A_1) A_3 \ ;\ \ \gamma_{29}=\frac{\dot{a}^4}{a^4N^4}A_1^2 (A_2-A_1)^2 \  ;\ \ \,\quad\gamma_{30}=\frac{\dot{a}^4}{a^4 N^4}A_1 A_2 (A_2-A_1)^2 \ ;\quad\quad\\
&\gamma_{31}=\frac{ \dot{a}^4}{a^4 N^4}A_2^2 (A_2-A_1)^2\,.
\end{align*}
We denote $\Gamma_i\equiv\frac{\delta}{\delta N}\Big[a^3N\,\gamma_i\Big]$ the constraint contributions from these basic expressions. All $\mathcal{E}$ and $\mathcal{F}$ terms can be expressed as linear combinations of the $\gamma$ terms, so their constraint equations will be equal to the same linear combination of the corresponding $\Gamma$ terms.

First we compute the contributions from all the $\gamma$ terms, and plug in them the no-boundary ansatz \eqref{ansatz}. Then we expand all $\Gamma$s to third order in $t$. Only nine out of these 31 terms actually start at order $t^{-1}$ (as we expected of terms where four covariant derivatives act on Riemann terms). They are, to leading order,\footnote{Beware that these equalities are only valid at order $t^{-1}$.}
\begin{align}
&\Gamma_{23}=\Gamma_{24}=\Gamma_{25}=-\frac{2a_3^2\big(a_3^2-a_1a_5\big)}{a_1^{13}t}\quad;\quad \Gamma_{26}=\Gamma_{27}=\Gamma_{28}=-\frac{2a_3^2\big(a_3^2-a_1a_5\big)}{3a_1^{13}t}\ ;\\
&\mbox{and}\quad \Gamma_{29}=\Gamma_{30}=\Gamma_{31}=-\frac{a_3^2\big(a_3^2-a_1a_5\big)}{3a_1^{13}t}\ .
\end{align}
In the $\mathcal{E}$ and $\mathcal{F}$ terms, these nine terms  appear in the eleven following combinations, which all give contributions that start at least at order $t$:
\begin{align}
&\frac{\delta}{\delta N}\bigg[\frac{\dot{a}^2}{a^2N^2}(A_2-A_1)^3A_4\bigg]\equiv\Gamma_{23}-2\Gamma_{24}+\Gamma_{25}=O(t^3)\ ;\\
&\frac{\delta}{\delta N}\bigg[\frac{\dot{a}^2}{a^2N^2}A_1(A_2-A_1)^2A_4\bigg]\equiv\Gamma_{24}-\Gamma_{23}=O(t)\ ;\\
&\frac{\delta}{\delta N}\bigg[\frac{\dot{a}^2}{a^2N^2}A_2(A_2-A_1)^2A_4\bigg]\equiv\Gamma_{25}-\Gamma_{24}=O(t)\ ;\\
&\frac{\delta}{\delta N}\bigg[\frac{\dot{a}^3}{a^3N^3}A_3(A_2-A_1)^3\bigg]=\Delta_4\equiv\Gamma_{26}-2\Gamma_{27}+\Gamma_{28}=O(t^3)\ ;\\
&\frac{\delta}{\delta N}\bigg[\frac{\dot{a}^3}{a^3N^3}A_3A_1(A_2-A_1)^2\bigg]\equiv\Gamma_{27}-\Gamma_{26}=O(t)\ ;\\
&\frac{\delta}{\delta N}\bigg[\frac{\dot{a}^3}{a^3N^3}A_3A_2(A_2-A_1)^2\bigg]\equiv\Gamma_{28}-\Gamma_{27}=O(t)\ ;\\
&\frac{\delta}{\delta N}\bigg[\frac{\dot{a}^4}{a^4N^4}(A_2-A_1)^4\bigg]=\Delta_2\equiv\Gamma_{29}-2\Gamma_{30}+\Gamma_{31}=O(t^3)\ ;\\
&\frac{\delta}{\delta N}\bigg[\frac{\dot{a}^4}{a^4N^4}A_1(A_2-A_1)^3\bigg]\equiv\Gamma_{30}-\Gamma_{29}=O(t)\ ;\\
&\frac{\delta}{\delta N}\bigg[\frac{\dot{a}^4}{a^4N^4}A_2(A_2-A_1)^3\bigg]\equiv\Gamma_{31}-\Gamma_{30}=O(t)\ ;\\
&\frac{\delta}{\delta N}\bigg[\frac{\dot{a}^3}{a^3N^3}A_1^2(A_2-A_1)\Big(A_3-\frac{\dot{a}}{aN}(A_2-A_1)\Big)\bigg]\equiv\Gamma_{26}-2\Gamma_{29}=O(t)\ ;\\
&\frac{\delta}{\delta N}\bigg[\frac{\dot{a}^2}{a^2N^2}A_1^2(A_2-A_1)\Big(A_4-\frac{\dot{a}}{aN}A_3-\frac{4\dot{a}^2}{a^2N^2}(A_2-A_1)\Big)\bigg]\equiv\Gamma_{23}-\Gamma_{26}-4\Gamma_{29}=O(t)\ .
\end{align} 
In fact, astonishingly, the cancellations go even further and the contribution at order $t$ also vanishes identically. The full expressions, which start at order $t^3,$ are listed in Appendix \ref{appendix:EandFconstraints}. Schematically, the order $t^3$ contribution of all $\delta\mathcal{F}$ terms can be written as
\begin{equation}
\delta\mathcal{F} =\frac{\big(a_3^2-a_1a_5\big)t^3}{a_1^{15}}\bigg[\lambda_1\big(a_3^2-a_1a_5\big)^2+\lambda_2a_1a_5\big(a_3^2-a_1a_5\big)+\lambda_3a_1a_3\big(a_3a_5-a_1a_7\big)\bigg]+O(t^5)\,;\label{ordert3F}
\end{equation}
where $\lambda_1$, $\lambda_2$  and $\lambda_3$ take different numerical values for each combination of derivatives.
As for the $\mathcal{E}$ terms, their contribution to the constraint is of the form
\begin{align}
\delta\mathcal{E}=\frac{t^3}{a_1^{15}}\bigg[&\mu_1\big(a_3^2-a_1a_5\big)^3+\mu_2a_1a_5\big(a_3^2-a_1a_5\big)^2+a_1a_3\big(\mu_3a_5a_3+\mu_4a_1a_7\big)\big(a_3^2-a_1a_5\big)\nonumber\\
&+\mu_5a_1^2a_3a_5\big(a_3a_5-a_1a_7\big)+\mu_6a_1^2a_3^2\big(a_1a_9-a_3a_7\big)\bigg]+O(t^5)\,;\label{ordert3E}
\end{align}
where $\mu_i$ are numerical factors varying for each case.

We are now in position to compute the type II string theory constraint equation up to fifth order in $\alpha^{\prime}$, and see whether this action admits a no-boundary solution.

%%%%%%%%

\subsection{Constraint equation for type II string theory\label{subsec:typeII}}

When compactified down to four dimensions, the type II action is of the form 
\begin{equation}
S_{\text{type II}}^{4d}=\frac{1}{2\kappa^2}\int\dd^4x\sqrt{-g}\Big[R-(\partial\phi)^2-2 V(\phi)+(\alpha^{\prime})^3\mathcal{E}_{(0,0)}\mathcal{R}^4+(\alpha^{\prime})^5\mathcal{E}_{(1,0)}\nabla^4\mathcal{R}^4+O(\alpha^{\prime\,6}) + \cdots \Big],\label{actionSTfinal}
\end{equation}
where we included a single scalar field with a potential $V(\phi),$ but where the ellipsis stands for many additional scalars and gauge fields, with the precise form of the action depending on the details of the compactification. In looking for no-boundary solutions we may once again neglect the contribution due to the gauge fields. In the same vein, the contributions in higher powers of $\alpha^\prime$ should be thought of as containing compactification dependent coefficient functions $\theta$, $\delta_i$, $\epsilon_i$ and $\eta_i,$ in front of the specific combinations $\mathcal{D,E,F}$ that we introduced in section \ref{subsec:d4R4}:
\begin{align}
\mathcal{E}_{(0,0)}\mathcal{R}^4=\theta \left( \mathcal{R}_1+\frac{246}{163}\mathcal{R}_2 \right)\quad\mbox{and}\quad\mathcal{E}_{(1,0)}\nabla^4\mathcal{R}^4=\sum_{i=1}^{4} \delta_i\mathcal{D}_i+\sum_{i=1}^{8}\epsilon_i\mathcal{E}_i+\sum_{i=1}^{8}\eta_i\mathcal{F}_i\,.
\end{align}

Does this theory now admit no-boundary solutions? As we demonstrated in the last section, the constraint equation, which provides the litmus test for the existence of regular solutions, does not receive $\alpha^\prime$ corrections at order $t^{-1}$ nor at order $t$ when the no-boundary ansatz \eqref{ansatz} is plugged in, due to the specific form of the ${\mathcal{D,E,F}}$ terms. This rather astonishing result may have an underlying explanation in the fact that no-boundary solutions approach Euclidean flat space smoothly near the South Pole, and hence covariant derivatives acting on the corresponding Riemann tensors are suppressed. In fact, the first non-trivial contributions to the constraint equation arise at order $t^3,$ where the constraint takes the form
\begin{align}
&-6a_3t^3-2V(\phi_0)a_1^3t^3+(\alpha^{\prime})^3\mathcal{E}_{(0,0)}\bigg[2205\cdot\frac{2a_3^2}{a_1^9}\big(3a_1a_5-4a_3^2\big)t^3-2934\cdot\frac{a_3^2}{a_1^9}\big(2a_3^2-a_1a_5\big)t^3\bigg]\nonumber\\
&+(\alpha^{\prime})^5\mathcal{E}_{(0,1)}\bigg[\#_1\cdot\frac{\big(a_3^2-a_1a_5\big)^3}{a_1^{15}}t^3+\#_2\cdot\frac{a_5\big(a_3^2-a_1a_5\big)^2}{a_1^{14}}+\big[\#_3\cdot a_3^2a_5+\#_4\cdot a_1a_3a_7\big]\frac{a_3^2-a_1a_5}{a_1^{14}}\nonumber\\
&\quad\quad\quad\quad\quad\quad+\#_5\cdot\frac{a_3a_5\big(a_3a_5-a_1a_7\big)}{a_1^{13}}+\#_6\cdot\frac{a_3^2\big(a_1a_9-a_3a_7\big)}{a_1^{13}}\bigg]=0\,.
\end{align}
Here we denoted $\phi(0)=\phi_0$ and the numerical coefficients at order $\alpha^{\prime 5}$ by $\#_i.$ In the absence of higher order corrections we would have learned that $\displaystyle a_3 = - \frac{V(\phi_0)}{3}a_1^3,$ i.e.~that the initial expansion rate depends on the location of the scalar field on the potential. Once the higher order terms are added, new families of solutions arise, and depending on the coefficient functions, $a_5, a_7$ and even $a_9$ can enter the constraint equation. At higher orders in $t,$ higher order terms in the series expansion for $a$ will of course also appear, and in this manner higher coefficients will continue to be given in terms of the lower order ones. Also, for terms with more derivatives, such as terms of the form $\nabla^6\mathcal{R}^4,$ we expect higher $a$ Taylor series coefficients to appear, in analogy with the results for $\mathcal{C}$ terms (see Appendix \ref{appendix:BandC}). For perturbative solutions, a self-consistency check will be that the solutions should have a smooth limit as $\alpha^\prime \to 0,$ very much like the limit $\beta \to 0$ encountered in section \ref{sec:qg} on quadratic gravity. What is clear however is that, given the current knowledge about $\alpha^\prime$ corrections, perturbative no-boundary solutions exist in type II string theory.

%%%%%%%%%%%%%%%%%%%%%%%%%%%%%%%%%%%%%%
%%%%%%%%%%%%%%%%%%%%%%%%%%%%%%%%%%%%%%

\section{Conclusions} \label{sec:conclusion}

The general expectation in cosmology is that as we approach the big bang, quantum gravity corrections will become more and more important, to the extent that we might remain ignorant about the initial stages of the universe until we will have fully uncovered quantum gravity. The no-boundary proposal, which is arguably the best understood theory for the initial conditions of the universe, goes somewhat against the grain by being formulated merely in semi-classical gravity. The question that concerned us in the present paper was whether the no-boundary proposal stands a chance of providing reliable answers given our current, partial, knowledge of quantum gravity. 

The very lack of a complete theory of quantum gravity means that we are not able to answer this question fully, yet the problem is still tractable to the extent that the general structure of perturbative quantum gravity corrections is known. Such corrections are expected to involve higher powers of the Riemann tensor as well as covariant derivatives acting on these tensors. The question thus becomes whether no-boundary solutions continue to exist in the presence of such correction terms. We have been able to derive explicit conditions, in particular Eq. \eqref{LOcondition}, that terms composed solely of Riemann tensors have to satisfy in order for no-boundary solutions to exist. This requirement is met for $f(R)$ gravity, quadratic gravity, Gauss-Bonnet gravity, heterotic string theory as well as type II string theory including the first non-trivial order in $\alpha^{\prime}$. What is more, by considering specific examples, we have been able to show that terms involving covariant derivatives acting on Riemann tensors may also coexist with no-boundary solutions. Here we studied the specific example provided by type II string theory up to order $\alpha^{\prime 5}.$ An interesting open question is whether the structure of string theory is such that it allows for no-boundary solutions in general.

Our results provide an important consistency check of the no-boundary proposal, as they show that for large classes of theories the results obtained in semi-classical gravity are robust. We should emphasise that our results apply both to inflationary and to ekpyrotic no-boundary instantons, these remaining the only classes of no-boundary instantons currently known. Our results in no way preclude the existence of qualitatively new solutions in full quantum gravity, but they do imply that no-boundary solutions will continue to exist in perturbative quantum gravity. Combined with the recent progress in constructing a consistent path integral implementation \cite{DiTucci:2019dji,DiTucci:2019bui,DiTucci:2020weq}, our results put the no-boundary proposal on a rather firm theoretical footing.

%%%%%%%%%%%%%%%%%%%%%%

\acknowledgments

We would like to thank Jan Gerken and Axel Kleinschmidt for useful discussions, as well as the anonymous referee for numerous suggestions, leading us in particular to include section \ref{sec:eft}. We gratefully acknowledge the support of the European Research Council in the form of the ERC Consolidator Grant CoG 772295 ``Qosmology''.

%%%%%%%%%%%%%%%%%%%%%%%%%%%%%%%%%%%%%%
%%%%%%%%%%%%%%%%%%%%%%%%%%%%%%%%

\appendix
\begin{small}
	
	\section{Constraint equation of Riemann terms in the no-boundary ansatz \label{appendix:constrainteq}}
	
		We plug the no-boundary ansatz \eqref{ansatz} into the Friedmann constraint equation \eqref{Friedmannconstraintnotsimplified}, and expand it at lowest orders in $t$. From \eqref{A1A2expansion} we know that at lowest order $A_1=A_2=\frac{a_3}{a_1N^2}$. Therefore we get
		\begin{align}
		0=2\pi^2\sum_{p_1,p_2}c_{p_1,p_2}&\bigg[2p_1(p_2-1)\frac{a_1t\cdot a_1^2}{N^2}\Big(\frac{a_3}{a_1N^2}\Big)^{P-1}+p_2(p_2-1)\frac{a_1t\cdot a_1\cdot a_3}{N^4}\Big(\frac{a_3}{a_1N^2}\Big)^{P-2}\nonumber\\
		&-p_2(2p_1+p_2-3)\frac{a_1t\cdot a_1^2}{N^2}\Big(\frac{a_3}{a_1N^2}\Big)^{P-1}\bigg]+O(t^3)\,;
		\end{align}
		where we defined $P=p_1+p_2$ for simplicity.\\
		This leading order equation can be further simplified to
		\begin{equation}
		2\pi^2\sum_{p_1,p_2}\frac{c_{p_1,p_2}}{N^{2P}}a_1^{4-P}a_3^{P-1}\left[2p_2-2p_1\right]t+O(t^3)=0\,.
		\end{equation}
		
		Let us now look at the next order. Because $a$ is an odd function of $t$, and hence $A_1$ and $A_2$ are even functions of $t$ (see \eqref{A1A2expansion}), the $t^2$ order of the Friedmann constraint will vanish. We directly consider the  $t^3$ order of the Friedmann constraint:
		\begin{align}
		2\pi^2&\sum_{p_1,p_2}c_{p_1,p_2}\bigg[2p_1(p_2-1)\frac{\big(a_1t+\frac{a_3t^3}{6}\big) \big(a_1+\frac{a_3t^2}{2}\big)^2}{N^2}\Big(\frac{a_3}{a_1N^2}+\frac{a_3^2-a_1a_5}{12N^4}t^2\Big)^{p_1-1}\Big(\frac{a_3}{a_1N^2}+\frac{a_3^2-a_1a_5}{6N^4}t^2\Big)^{p_2}\nonumber\\
		&+p_2(p_2-1)\frac{\big(a_1t+\frac{a_3t^3}{6}\big)\big(a_1+\frac{a_3t^2}{2}\big) \big(a_3+\frac{a_5t^2}{2}\big)}{N^4}\Big(\frac{a_3}{a_1N^2}+\frac{a_3^2-a_1a_5}{12N^4}t^2\Big)^{p_1}\Big(\frac{a_3}{a_1N^2}+\frac{a_3^2-a_1a_5}{6N^4}t^2\Big)^{p_2-2}\nonumber\\
		&-p_2(2p_1+p_2-3)\frac{\big(a_1t+\frac{a_3t^3}{6}\big)\cdot \big(a_1+\frac{a_3t^2}{2}\big)^2}{N^2}\Big(\frac{a_3}{a_1N^2}+\frac{a_3^2-a_1a_5}{12N^4}t^2\Big)^{p_1}\Big(\frac{a_3}{a_1N^2}+\frac{a_3^2-a_1a_5}{6N^4}t^2\Big)^{p_2-1}\nonumber\\
		&+(1-p_2)\big(a_1t\big)^3\Big(\frac{a_3}{a_1N^2}\Big)^{P}\bigg]+O(t^5)=0\,.
		\end{align}
		This can then be simplified to
		\begin{align}
		&2\pi^2\sum_{p_1,p_2}\frac{c_{p_1,p_2}}{N^{2P}}a_1^{3-P}a_3^{P-2}\Big(a_3^2\cdot G_3[p_1,p_2]+a_1a_5\cdot G_{5}[p_1,p_2]\Big)t^3+O(t^5)=0\,,
		\end{align}
		with
		\begin{align}
		G_3[p_1,p_2]=\frac{1}{6}\left(p_1^2-15p_1+6-4p_2^2+12p_2\right)\ \mbox{and}\ 
		G_5[p_1,p_2]=\frac{p_1(1-p_1)}{6}-\frac{2p_2(1-p_2)}{3}\,.
		\end{align}
	
	\section{Constraint equations for $\mathcal{B}$ and $\mathcal{C}$ terms\label{appendix:BandC}}
	Here we display the constraint equations of $\mathcal{B}$ terms where the no-boundary ansatz has been plugged in.
	Writing $\delta\mathcal{B}\equiv\frac{\delta}{\delta N}\big(a^3N\mathcal{B}\big),$ we find
	\begin{align}
	\delta\mathcal{B}_1=\,&-12\bigg[\frac{4}{15a_1^6}\Big(25a_3^3-29a_1a_3a_5+4a_1^2a_7\Big)\, t^3\bigg]+O(t^5)\,;\\
	\delta\mathcal{B}_2=&-12\bigg[\frac{1}{15a_1^6}\Big(85a_3^3-101a_1a_3a_5+16a_1^2a_7\Big)\, t^3\bigg]+O(t^5)\,;\quad\quad\\
	\delta\mathcal{B}_3=&-36\bigg[\frac{2}{15a_1^6}\Big(35a_3^3-43a_1a_3a_5+8a_1^2a_7\Big)\,t^3\bigg]+O(t^5)\,;\\
	\delta\mathcal{B}_4=\,&-\frac{4}{5a_1^6}\Big(35a_3^3-43a_1a_3a_5+8a_1^2a_7\Big)t^3+O(t^5)\,.
	\end{align}
	All those $\nabla^2R^2$ terms possess a no-boundary solution which specifies $a_7$ in terms of $a_1$, $a_3$, and $a_5$, but where the latter are not specified by the $\nabla^2R^2$ terms alone.
	
	We now look at the constraint equations for $\mathcal{C}$ terms. Their expressions in terms of $A_1$, $A_2$, $A_3$ and $A_4$ are
	\begin{align}
	\mathcal{C}_1&=12\bigg[4\Big[A_2(A_2-A_1)+\frac{2\dot{a}^2}{a^2N^2}(A_2-A_1)+\frac{\dot{a}}{aN}A_3\Big]^2+\Big[A_4-\frac{4\dot{a}^2}{a^2N^2}(A_2-A_1)+\frac{\dot{a}}{aN}A_3\Big]^2\bigg]\,;\label{nabla4R2initial}\\
	\mathcal{C}_2&=\ 12\bigg[A_4^2+\frac{4\dot{a}}{aN}A_3A_4+\frac{7\dot{a}^2}{a^2N^2}A_3^2+2A_4A_2(A_2-A_1)+4A_2^2(A_2-A_1)^2-\frac{4\dot{a}^2}{a^2N^2}A_4(A_2-A_1)\nonumber\\
	&+\frac{16\dot{a}^4}{a^4N^4}(A_2-A_1)^2+\frac{8\dot{a}^2}{a^2N^2}A_2(A_2-A_1)^2+\frac{10\dot{a}}{aN}A_3A_2(A_2-A_1)+\frac{4\dot{a}^3}{a^3N^3}A_3(A_2-A_1)\bigg]\, ;\\
	\mathcal{C}_3&=36\bigg[\frac{3\dot{a}}{aN}A_3+2A_2(A_2-A_1)+A_4\bigg]^2\,;\\
	\mathcal{C}_4&
	=12\bigg[A_4^2-\frac{4\dot{a}}{aN}A_3A_4+\frac{19\dot{a}^2}{a^2N^2}A_3^2+\frac{16\dot{a}}{aN}A_3A_2(A_2-A_1)-\frac{80\dot{a}^3}{a^3N^3}A_3(A_2-A_1)\nonumber\\
	&\quad+\frac{160\dot{a}^4}{a^4N^4}(A_2-A_1)^2-\frac{48\dot{a}^2}{a^2N^2}A_2(A_2-A_1)^2+8A_2^2(A_2-A_1)^2\bigg]\,;\\
	\mathcal{C}_5&=12\bigg[A_4^2-\frac{2\dot{a}}{aN}A_3A_4+\frac{11\dot{a}^2}{a^2N^2}A_3^2-\frac{34\dot{a}^3}{a^3N^3}A_3\big(A_2-A_1\big)+\frac{8\dot{a}}{aN}A_3A_2\big(A_2-A_1\big)+2A_4A_2\big(A_2-A_1\big)\nonumber\\
	&\quad-\frac{6\dot{a}^2}{a^2N^2}A_4\big(A_2-A_1\big)+\frac{104\dot{a}^4}{a^4N^4}\big(A_2-A_1\big)^2-\frac{36\dot{a}^2}{a^2N^2}A_2\big(A_2-A_1\big)^2+6A_2^2\big(A_2-A_1\big)^2\bigg]\,;\\
	\mathcal{C}_6&=36 \bigg[\Big[A_4+2A_2(A_2-A_1)\Big]^2-\frac{12\dot{a}^2}{a^2N^2}(A_2-A_1)
	A_4-\frac{24\dot{a}^2}{a^2N^2} A_2(A_2-A_1)^2+\frac{3\dot{a}^2}{a^2N^2}A_3^2\nonumber\\
	&\quad+\frac{12\dot{a}^3}{a^3N^3}(A_2-A_1) A_3+\frac{48\dot{a}^4}{a^4N^4} (A_2-A_1)^2\bigg]\,.\label{nabla4R2final}
	\end{align}
	Writing $\delta\mathcal{C}\equiv\frac{\delta}{\delta N} \big(a^3N\mathcal{C}\big),$ we find
	\begin{align}
	\delta\mathcal{C}_1&=\frac{8 t^3 \left(3262 a_1 a_3^2
		a_5+60 a_1^3 a_9-2135 a_3^4-a_1^2
		\left(592 a_3 a_7+595
		a_5^2\right)\right)}{35 a_1^9}+O\left(t^5\right)\,;\\
	\delta\mathcal{C}_2&=\frac{4 t^3 \left(3528 a_1 a_3^2
		a_5-5 a_1^2 \left(161a_5^2-24a_1a_9\right)-1995 a_3^4-848a_1^2
		a_3 a_7\right)}{35a_1^9}+O\left(t^5\right)\,;\\
	\delta\mathcal{C}_3&=\frac{48 t^3 \left(133 a_1 a_3^2
		a_5-15 a_1^2 \left(7 a_5^2-2
		a_1 a_9\right)+70 a_3^4+128
		a_3 a_7 N^2\right)}{35
		a_1^9}+O\left(t^5\right)\,;\\
	\delta\mathcal{C}_4&=\frac{8 t^3 \left(1008 a_1 a_3^2
		a_5+12 a_1^3 a_9-735 a_3^4-19 a_1^2
		\left(8 a_3 a_7+7
		a_5^2\right)\right)}{7 a_1
		N^8}+O\left(t^5\right)\,;\\
	\delta\mathcal{C}_5&=\frac{2 t^3 \left(13048 a_1 a_3^2
		a_5-5 a_1^2 \left(413
		a_5^2-48 a_1 a_9\right)-8855
		a_3^4-2368 a_1^2a_3 a_7\right)}{35 a_1^9}+O\left(t^5\right)\,;\\
	\delta\mathcal{C}_6&=\frac{12 t^3 \left(2968 a_1 a_3^2
		a_5-15a_1^2 \left(49
		a_5^2-8 a_1 a_9\right)-1505
		a_3^4-848a_1^2 a_3 a_7\right)}{35 a_1^9}+O\left(t^5\right)\,.
	\end{align}
	
	These six $\nabla^4R^2$ terms all admit a regular no-boundary solution, for which the coefficient $a_9$ is fixed in terms of $a_1$, $a_3$, $a_5$ and $a_7$ at order $t^3$ of the constraint. This ensures the existence of a solution if these $\nabla^4R^2$ terms are combined with Riemann terms and $\nabla^2R^2$ terms, since $a_9$ is a new degree of freedom at order $t^3$.
	\section{Constraint equations from $\mathcal{D}$, $\mathcal{E}$ and $\mathcal{F}$ terms\label{appendix:EandFconstraints}}
	Expressions of $\mathcal{D}$ terms as functions of $A_1$, $A_2$ and $A_3$:
	\begin{align}
	&\mathcal{D}_1=144\bigg[A_3^4+\frac{16\dot{a}^2}{a^2N^2}A_3^2\big(A_2-A_1\big)^2+\frac{64\dot{a}^4}{a^4N^4}\big(A_2-A_1\big)^4\bigg]\,;\quad\quad\\
	&\mathcal{D}_2=\,48\bigg[3A_3^4+\frac{24\dot{a}^2}{a^2N^2}A_3^2(A_2-A_1)^2+\frac{64\dot{a}^4}{a^4N^4}\big(A_2-A_1)^4\bigg]\,;\quad\\
	&\mathcal{D}_3=\,48\bigg[A_3^4+\frac{4\dot{a}^2}{a^2N^2}A_3^2(A_2-A_1)^2+\frac{40\dot{a}^4}{a^4N^4}\big(A_2-A_1\big)^4\bigg]\,;\\
	&\mathcal{D}_4=\,48\bigg[A_3^4+\frac{16\dot{a}^3}{a^3N^3}A_3 \big(A_2-A_1\big)^3+\frac{20\dot{a}^4}{a^4N^4} \big(A_2-A_1\big)^4\bigg]\,.\quad
	\end{align}
	Expressions of $\mathcal{E}$ terms as functions of $A_1$, $A_2$, $A_3$ and $A_4$:
	\begin{align}
	\mathcal{E}_1=&144\big(A_1^2+A_2^2\big)\Bigg[4 \Big(\frac{\dot{a}}{aN}A_3+\frac{2\dot{a}^2}{a^2N^2}(A_2-A_1)+A_2(A_2-A_1)\Big)^2\nonumber\\
	&\quad\quad\quad\quad\quad\quad+\Big(A_4+\frac{4\dot{a}^2}{a^2N^2}(A_1-A_2)+\frac{\dot{a}}{aN}A_3\Big)^2\Bigg]\,;\\
	\mathcal{E}_2=&144\big(A_1^2+A_2^2\big)\Bigg[A_4^2-\frac{4\dot{a}}{aN}A_4A_3+\frac{19\dot{a}^2}{a^2N^2}A_3^2+\frac{16\dot{a}}{aN}A_3A_2(A_2-A_1)+\frac{80\dot{a}^3}{a^3N^3}A_3(A_1-A_2)\nonumber\\
	&\quad\quad\quad\quad\quad\quad+8A_2^2(A_2-A_1)^2+\frac{160\dot{a}^4}{a^4N^4}(A_2-A_1)^2-\frac{48\dot{a}^2}{a^2N^2}A_2(A_2-A_1)^2\Bigg]\,;\\
	\mathcal{E}_3=&144\Bigg[A_2A_4+\frac{\dot{a}}{aN}A_3(A_2+2A_1)-2A_1A_2(A_1-A_2)-\frac{4\dot{a}^2}{a^2N^2}(A_1-A_2)^2\Bigg]^2
	\,;\\
	\mathcal{E}_4=&48\Bigg[12A_1^2A_2^2(A_2-A_1)^2+3A_2^2A_4^2+12A_1A_2^2(A_2-A_1)A_4\nonumber\\
	&\quad+\frac{16\dot{a}^4}{a^4N^4}(A_2-A_1)^2(A_2^2-5A_1A_2+13A_1^2)-\frac{12\dot{a}}{aN}A_3A_2(A_2-A_1)\Big(A_4+2A_1(A_2-A_1)\Big)\nonumber\\
	&\quad-\frac{12\dot{a}^3}{a^3N^3}A_3(A_2-A_1)\Big(2(A_2-A_1)^2-4A_1(A_2-A_1)-3A_1A_2\Big)\nonumber\\
	&\quad-\frac{12\dot{a}^2}{a^2N^2}(A_2-A_1)\Big(3A_1A_2A_4-A_3^2(A_2-A_1)+6A_1^2A_2(A_2-A_1)\Big)+\frac{9\dot{a}^2}{a^2N^2}A_3^2A_2^2\Bigg]\,;\\
	\mathcal{E}_5=&48\Bigg[4A_1^2 \Big[\frac{\dot{a}}{aN}A_3+\frac{2\dot{a}^2}{a^2N^2}(A_2-A_1)+A_2(A_2-A_1)\Big]^2+A_2^2\Big[A_4+\frac{\dot{a}}{aN}A_3+\frac{4\dot{a}^2}{a^2N^2}(A_1-A_2)\Big]^2\Bigg]\,;\\
	\mathcal{E}_6=&48\Bigg[16\frac{\dot{a}^4}{a^4N^4}(A_2-A_1)^2(7A_1^2+3A_2^2)-12\frac{\dot{a}^2}{a^2N^2}A_2(A_2-A_1)^2(A_2^2+3A_1^2)\nonumber\\
	&\quad+2A_2^2(A_2-A_1)^2(A_2^2+3A_1^2)+\frac{\dot{a}^2}{a^2N^2}A_3^2(8A_1^2+11A_2^2)+16A_3\frac{\dot{a}^3}{a^3N^3}(A_1-A_2)(2A_2^2+3A_1^2)\nonumber\\
	&\quad+4\frac{\dot{a}}{aN}A_3A_2(A_2-A_1)(A_2^2+3A_1^2)+A_2^2A_4^2-4\frac{\dot{a}}{aN}A_3A_2^2A_4\Bigg]\,;\\
	\mathcal{E}_7=&48\Bigg[\frac{16\dot{a}^4}{a^4N^4}(A_2-A_1)^2(A_1^2+A_2^2)+4A_1^2A_2^2(A_2-A_1)^2-\frac{8\dot{a}^3}{a^3N^3}A_3(A_2-A_1)(A_2^2-2A_1^2)\nonumber\\
	&\quad+\frac{\dot{a}^2}{a^2N^2}(4A_1^2+A_2^2)A_3^2+\frac{8\dot{a}^2}{a^2N^2}A_2^2A_4(A_1-A_2)+\frac{16\dot{a}^2}{a^2N^2}A_1^2A_2(A_2-A_1)^2\nonumber\\
	&\quad+\frac{2\dot{a}}{aN}A_2^2A_3A_4+\frac{8\dot{a}}{aN}A_1^2A_2A_3(A_2-A_1)+A_2^2A_4^2\Bigg]=\mathcal{E}_5\ ;\\
	\mathcal{E}_8=&48\Bigg[\frac{4\dot{a}^4}{a^4N^4}(A_2-A_1)^2(3A_2^2+18A_2A_1+19A_1^2)-\frac{8\dot{a}^3}{a^3N^3}A_3(A_2-A_1)(A_2^2+6A_2A_1+3A_1^2)\nonumber\\
	&\quad+\frac{\dot{a}^2}{a^2N^2}A_3^2(8A_2A_1+7A_2^2+4A_1^2)-\frac{24\dot{a}^2}{a^2N^2}A_1A_2(A_2-A_1)^2(A_2+A_1)\nonumber\\
	&\quad+\frac{8\dot{a}}{aN}A_2A_1A_3(A_2^2-A_1^2)-\frac{4\dot{a}}{aN}A_2^2A_3A_4+4A_2^2A_1(A_2-A_1)^2(A_2+A_1)+A_2^2A_4^2\Bigg]\,.
	\end{align}
	
	Expressions of $\mathcal{F}$ terms through $A_1$, $A_2$, $A_3$ and $A_4$ quantities:
	\begin{align}
	\mathcal{F}_1=&144\bigg[\frac{8\dot{a}^2}{a^2N^2}(A_2-A_1)^2+A_3^2\bigg]\bigg[2 A_2A_1(A_2-A_1)-\frac{4\dot{a}^2}{a^2N^2}(A_2-A_1)^2+\frac{\dot{a}}{aN}A_3(A_2+2A_1)+A_2 A_4\bigg]\,;\\
	\mathcal{F}_2=&48\bigg[-\frac{16\dot{a}^4}{a^4N^4}(A_2-A_1)^3(A_2+2A_1)-\frac{12\dot{a}^3}{a^3N^3}A_3(A_2-A_1)^2(A_2-2A_1)\nonumber\\
	&\quad+\frac{24\dot{a}^2}{a^2N^2}A_1A_2(A_2-A_1)^3+\frac{18\dot{a}^2}{a^2N^2}A_3^2A_1(A_1-A_2)+\frac{12\dot{a}^2}{a^2N^2}A_2A_4(A_2-A_1)^2\nonumber\\
	&\quad+\frac{6\dot{a}}{aN}A_3^3(A_1-A_2)+6A_3^2A_1A_2(A_2-A_1)+3A_2A_3^2A_4\bigg]\,;\\
	\mathcal{F}_3=&144 \bigg[\frac{2\dot{a}}{aN}A_1(A_2-A_1)+A_2 A_3\bigg]\bigg[\frac{\dot{a}}{aN}A_3^2+\frac{4\dot{a}}{aN}A_2(A_2-A_1)^2+ \frac{8\dot{a}^3}{a^3N^3}(A_2-A_1)^2+A_3 A_4\bigg]\,;\\
	\mathcal{F}_4=&144\bigg[\frac{2\dot{a}}{aN}A_1(A_2-A_1)+A_2 A_3\bigg] \bigg[\frac{4\dot{a}}{aN}(A_2-A_1)^2\big(A_2-\frac{6\dot{a}^2}{a^2N^2}\big)-\frac{2\dot{a}}{aN}A_3^2+\frac{8\dot{a}^2}{a^2N^2}A_3(A_2-A_1)+A_3 A_4\bigg]\,;\\
	\mathcal{F}_5=&48\bigg[-\frac{8\dot{a}^4}{a^4N^4}(A_2-A_1)^3(A_2-3A_1)+\frac{2\dot{a}^3}{a^3N^3}A_3(A_2-A_1)^2(A_2+6A_1)-\frac{4\dot{a}^2}{a^2N^2}A_2A_3^2(A_2-A_1)\nonumber\\
	&\quad+\frac{2\dot{a}^2}{a^2N^2}A_2A_4(A_2-A_1)^2+\frac{12\dot{a}^2}{a^2N^2}A_1A_2(A_2-A_1)^3+\frac{\dot{a}}{aN}A_2A_3^3+A_2A_3^2A_4\bigg]\,;\\
	\mathcal{F}_6=&48\bigg[-\frac{2\dot{a}^4}{a^4N^4}(A_2-A_1)^3(9A_2+13A_1)+\frac{6\dot{a}^3}{a^3N^3}A_1A_3(A_2-A_1)^2+\frac{2\dot{a}^2}{a^2N^2}A_3^2(A_2^2-A_1^2)\nonumber\\
	&\quad+\frac{12\dot{a}^2}{a^2N^2}A_1A_2(A_2-A_1)^3+\frac{2\dot{a}}{aN}A_2^2A_3(A_2-A_1)^2-\frac{2\dot{a}}{aN}A_2A_3^3+A_2A_3^2A_4\bigg]\,;\\
	\mathcal{F}_7=&48\bigg[\frac{8\dot{a}^4}{a^4N^4}(A_2-A_1)^3(A_2+A_1)+\frac{2\dot{a}^3}{a^3N^3}A_3(A_2-A_1)^2(2A_2+5A_1)+\frac{2\dot{a}^2}{a^2N^2}A_1A_4(A_2-A_1)^2\nonumber\\
	&\quad-\frac{4\dot{a}^2}{a^2N^2}A_2A_3^2(A_2-A_1)+\frac{4\dot{a}^2}{a^2N^2}A_2(A_2-A_1)^3(A_2+2A_1)+\frac{\dot{a}}{aN}A_2A_3^3+A_2A_3^2A_4\bigg]\,;\\
	\mathcal{F}_8=&48\bigg[\frac{2\dot{a}^4}{a^4N^4}(A_2-A_1)^3(5A_2-27A_1)-\frac{6\dot{a}^3}{a^3N^3}A_3(A_2-A_1)^2(2A_2-3A_1)+\frac{12\dot{a}^2}{a^2N^2}A_1A_2(A_2-A_1)^3\nonumber\\
	&\quad+\frac{4\dot{a}^2}{a^2N^2}A_2A_3^2(A_2-A_1)+\frac{2\dot{a}}{aN}A_2^2A_3(A_2-A_1)^2-\frac{2\dot{a}}{aN}A_2A_3^3+A_2A_3^2A_4\bigg]\,.
	\end{align}

	We display here the constraint equations obtained for all $\mathcal{D}$, $\mathcal{E}$ and $\mathcal{F}$ terms (using again the notation $\delta\mathcal{D}\equiv\frac{\delta}{\delta N} \big(a^3N\mathcal{D}\big)$ and similarly for $\mathcal {E}$ and $\mathcal {F}$):
	\begin{align}
	\delta\mathcal{D}_1&=144\Big[\Delta_1+16\Delta_3+64\Delta_2\Big]=\frac{384(a_3^2-a_1a_5)^3}{a_1^{15}}\,t^3+O\left(t^5\right)\,;\\
	\delta\mathcal{D}_2&=\,12\Big[64\Delta_2+12\big[\Delta_1+8\Delta_3+16\Delta_2\big]\Big]=\frac{1408(a_3^2-a_1a_5)^3}{3
		a_1^{15}}\,t^3+O\left(t^5\right)\,;\\
	\delta\mathcal{D}_3&=\,2\Big[864\Delta_2+24\big[\Delta_1+4\Delta_3+4\Delta_2\big]\Big]=\frac{144(a_3^2-a_1a_5)^3}{
		a_1^{15}}\, t^3+O\left(t^5\right)\,;\\
	\delta\mathcal{D}_4&=\,4\Big[240\Delta_2+192\Delta_4+12\Delta_1\Big]=\frac{136
		(a_3^2-a_1a_5)^3}{a_1^{15}}\, t^3+O\left(t^5\right)\,;\\
	\delta {\mathcal E}_1&=144\cdot\frac{4 t^3}{105a_1^{15}}\bigg[60 a_1^2 a_3^2 \big(a_1a_9-a_3a_7\big)+532a_1^2a_3a_5(a_3a_5-a_1a_7\big)-3675 \big(a_3^2-a_1a_5\big)^3\nonumber\\
	&\quad-3535a_1a_5\big(a_3^2-a_1a_5\big)^2-a_1a_3\big(616a_5a_3+924a_1a_7\big)\big(a_3^2-a_1a_5\big)\bigg]+O\left(t^5\right)\,;\\
	\delta {\mathcal E}_2&=144\cdot\frac{4 t^3}{63a_1^{15}}\bigg[36 a_1^2 a_3^2 \big(a_1a_9-a_3a_7\big)+420a_1^2a_3a_5(a_3a_5-a_1a_7\big)-2793 \big(a_3^2-a_1a_5\big)^3\nonumber\\
	&\quad-2765a_1a_5\big(a_3^2-a_1a_5\big)^2-a_1a_3\big(1246a_5a_3+560a_1a_7\big)\big(a_3^2-a_1a_5\big)\bigg]+O\left(t^5\right)\,;\\
	\delta {\mathcal E}_3&=144\cdot\frac{2 t^3}{315a_1^{15}}\bigg[180 a_1^2 a_3^2 \big(a_1a_9-a_3a_7\big)+588a_1^2a_3a_5(a_3a_5-a_1a_7\big)-6860 \big(a_3^2-a_1a_5\big)^3\nonumber\\
	&\quad-5915a_1a_5\big(a_3^2-a_1a_5\big)^2-a_1a_3\big(-4046a_5a_3+2996a_1a_7\big)\big(a_3^2-a_1a_5\big)\bigg]+O\left(t^5\right)\,;\\
	\delta {\mathcal E}_4&=48\cdot\frac{2 t^3}{315a_1^{15}}\bigg[540a_1^2 a_3^2 \big(a_1a_9-a_3a_7\big)+3276a_1^2a_3a_5(a_3a_5-a_1a_7\big)-23450 \big(a_3^2-a_1a_5\big)^3\nonumber\\
	&\quad-22575a_1a_5\big(a_3^2-a_1a_5\big)^2-a_1a_3\big(-4179a_5a_3+7644a_1a_7\big)\big(a_3^2-a_1a_5\big)\bigg]+O\left(t^5\right)\,;\\
	\delta {\mathcal E}_5&=\delta E_7=48\cdot\frac{t^3}{315a_1^{15}}\bigg[360 a_1^2 a_3^2 \big(a_1a_9-a_3a_7\big)+3192a_1^2a_3a_5(a_3a_5-a_1a_7\big)-29470 \big(a_3^2-a_1a_5\big)^3\nonumber\\
	&\quad-28000a_1a_5\big(a_3^2-a_1a_5\big)^2-a_1a_3\big(2240a_5a_3+7000a_1a_7\big)\big(a_3^2-a_1a_5\big)\bigg]+O\left(t^5\right)\,;\\
	\delta {\mathcal E}_6&=48\cdot\frac{t^3}{315a_1^{15}}\bigg[360 a_1^2 a_3^2 \big(a_1a_9-a_3a_7\big)+4200a_1^2a_3a_5(a_3a_5-a_1a_7\big)-32795 \big(a_3^2-a_1a_5\big)^3\nonumber\\
	&\quad-32305a_1a_5\big(a_3^2-a_1a_5\big)^2-a_1a_3\big(11396a_5a_3+6664a_1a_7\big)\big(a_3^2-a_1a_5\big)\bigg]+O\left(t^5\right)\,;\\
	\delta {\mathcal E}_8&=48\cdot\frac{t^3}{630a_1^{15}}\bigg[720 a_1^2 a_3^2 \big(a_1a_9-a_3a_7\big)+8400a_1^2a_3a_5(a_3a_5-a_1a_7\big)-65695 \big(a_3^2-a_1a_5\big)^3\nonumber\\
	&\quad-64610a_1a_5\big(a_3^2-a_1a_5\big)^2-a_1a_3\big(22792a_5a_3+13328a_1a_7\big)\big(a_3^2-a_1a_5\big)\bigg]+O\left(t^5\right)\,;\\
	\delta {\mathcal F}_1&=144\cdot\frac{2\big(a_3^2-a_1a_5\big) t^3}{45a_1^{15}}\bigg[235a_3^2(a_3^2-a_1a_5)-64a_1a_3\big(a_3a_5-a_1a_7\big)\bigg]+O\left(t^5\right)\,;\\
	\delta {\mathcal F}_2&=48\cdot\frac{2\big(a_3^2-a_1a_5\big) t^3}{45a_1^{15}}\bigg[105a_1a_5\big(a_3^2-a_1a_5\big)+12a_1a_3\big(a_3a_5-a_1a_7\big)-160\big( a_3^2-a_1a_5\big)^2\bigg]+O\left(t^5\right)\,;\\
	\delta {\mathcal F}_3&=144\cdot\frac{4\big(a_3^2-a_1a_5\big) t^3}{45a_1^{15}}\bigg[60a_1a_5\big(a_3^2-a_1a_5\big)-18a_1a_3\big(a_3a_5-a_1a_7\big)+5\big(a_3^2-a_1a_5\big)^2\bigg]+O\left(t^5\right)\,;\\
	\delta {\mathcal F}_4&=144\cdot\frac{2\big(a_3^2-a_1a_5\big) t^3}{45a_1^{15}}\bigg[-155a_1a_5\big(a_3^2-a_1a_5\big)+32a_1a_3\big(a_3a_5-a_1a_7\big)-30\big(a_3^2-a_1a_5\big)^2\bigg]+O\left(t^5\right)\,;\\
	\delta {\mathcal F}_5&=48\cdot\frac{ \big(a_3^2-a_1a_5\big)t^3}{90a_1^{15}}\bigg[725a_3^2(a_3^2-a_1a_5)-128a_1a_3\big(a_3a_5-a_1a_7\big)\bigg]+O\left(t^5\right)\,;\\
	\delta {\mathcal F}_6&=48\cdot\frac{\big(a_3^2-a_1a_5\big)t^3}{180a_1^{15}}\bigg[-725a_1a_5\big(a_3^2-a_1a_5\big)+128a_1a_3\big(a_3a_5-a_1a_7\big)-265\big(a_3^2-a_1a_5\big)^2\bigg]+O\left(t^5\right)\,;\\
	\delta {\mathcal F}_7&=48\cdot\frac{\big(a_3^2-a_1a_5\big)t^3}{90a_1^{15}}\bigg[725a_1a_5\big(a_3^2-a_1a_5\big)-128a_1a_3\big(a_3a_5-a_1a_7\big)+10\big(a_3^2-a_1a_5\big)^2\bigg]+O\left(t^5\right)\,;\\
	\delta {\mathcal F}_8&=48\cdot\frac{\big(a_3^2-a_1a_5\big)t^3}{180a_1^{15}}\bigg[-725a_1a_5\big(a_3^2-a_1a_5\big)+128a_1a_3\big(a_3a_5-a_1a_7\big)-275\big(a_3^2-a_1a_5\big)^2\bigg]+O\left(t^5\right).
	\end{align}
\end{small}

\bibliographystyle{utphys}
\bibliography{NoBoundarySolutionQGCorrection}

\end{document}